\documentclass[a4paper,11pt]{article}
\usepackage{jcappub} % for details on the use of the package, please
\usepackage[normalem]{ulem}
\usepackage{multirow}
\makeatletter
\gdef\@fpheader{ }
\makeatother
\hypersetup{
	colorlinks   = true, %Colours links instead of ugly boxes
	urlcolor     = blue, %Colour for external hyperlinks
	linkcolor    = blue, %Colour of internal links
	citecolor    = red   %Colour of citations
}

\newcommand{\N}{{N}}

\title{Tail diversity from inflation}

\author[a]{S. Hooshangi,}
\emailAdd{sina.hooshangi@ipm.ir}

\author[a]{M. H. Namjoo,}
\emailAdd{mh.namjoo@ipm.ir}

\author[b,a]{and  M. Noorbala}
\emailAdd{mnoorbala@ut.ac.ir}
\affiliation[a]{School of Astronomy, Institute for Research in Fundamental Sciences (IPM), Tehran, Iran, P.O. Box 19395-5531}
\affiliation[b]{Department of Physics, University of Tehran, Iran. P.O.\ Box 14395-547}

\abstract{The tail of the distribution of primordial fluctuations (corresponding to the likelihood of realization of large fluctuations) is of interest, from both theoretical and observational perspectives. In particular, it is relevant for the accurate evaluation of the primordial black hole (PBH) abundance. In this paper, we first analyze the non-perturbative $\delta N$ formalism as a method to non-perturbatively estimate the probability distribution function (PDF)  of primordial fluctuations, discuss its underlying assumptions and deal with several subtleties that may arise as a result of considering large fluctuations. Next, we employ the method to study several non-attractor single-field inflationary models as the simplest examples that may lead to the abundant production of PBHs. We conclude that the Gaussian extrapolation from linear perturbation theory may fail drastically to predict the likelihood of large fluctuations. Specifically, we show that a truncation of the tail, a power-law tail, a double-exponential tail, and a doubly peaked distribution can all be realized for the curvature perturbation in the single-field non-attractor models of inflation. We thus show that there is a diverse zoo of possible tails from inflation so that a model-dependent, non-perturbative study of the distribution of the primordial fluctuations seems inevitable concerning PBH abundance. 
}

\begin{document}
	
	\maketitle
	
	% check L3/L2 and \cal S for recent potentials
	%\tableofcontents
	\section{Introduction} 
	
	The LIGO/Virgo~\cite{LIGOScientific:2016aoc} observations of binary mergers have revived interest in primordial black holes (PBHs) as a possible source for the observed gravitational waves. PBHs may be formed as a result of the generation of sufficiently large fluctuations during inflation which then collapse upon horizon re-entry in the radiation-dominated era~\cite{Carr:1974nx, Carr:1975qj}. The abundance (and therefore the merger rate) of PBHs is thus related to the statistics of primordial fluctuations. The realization of large fluctuations---namely, ${\cal O}(1)$ density contrast---which corresponds to sampling from the tail of the probability distribution functions (PDFs) casts doubts on the validity of the perturbation theory in two ways. First, the perturbation theory is clearly expected to hold only for small fluctuations. Second, while a Gaussian approximation can be sensible for describing the peak of the distribution---as predicted by the linear perturbation theory---it is not as justified for tracing the tail. Therefore, non-perturbative methods must be employed for the analysis of the tail of the distributions and the accurate estimation of the PBH number density. A number of non-perturbative methods exist in the related literature.  Ref.~\cite{Celoria:2021vjw} applied the non-perturbative wave-function method to estimate the tail of the distribution for a particular type of interaction.  Ref.~\cite{Ezquiaga:2019ftu} used the stochastic formalism~\cite{Vennin:2015hra} which is also believed to hold non-perturbatively as long as its basic assumptions hold.    In Ref.~\cite{Hooshangi:2021ubn} the so-called $\delta N$ formalism in its non-perturbative form has been employed for studying the behavior of the PDF at its tail. Working with the curvature perturbation on uniform hypersurfaces, $\zeta$, it is remarkable that in all aforementioned literature, the tails of the distributions in the considered models are shown to be significantly different from the naive, perturbative expectations. In particular, Ref.~\cite{Ezquiaga:2019ftu} shows that an exponential tail can be developed through the stochastic effects ( with a model-dependent  exponent~\cite{Pattison:2017mbe, Ballesteros:2020sre, Pattison:2021oen, Figueroa:2020jkf, Figueroa:2021zah, Ahmadi:2022lsm}) while Ref.~\cite{Hooshangi:2021ubn} shows that even a significantly heavier (namely, a power-law) tail is also possible in certain, non-attractor models of inflation.  A truncation of the tail, i.e., total suppression of the probability density, is reported in Refs.~\cite{Cai:2021zsp, Cai:2022erk}.
	
	Given the significance of the accurate estimation of the PBH abundance, it is thus important to make clear the assumptions under which each non-perturbative method is applicable. Exploring the generality of a non-trivial tail in inflationary models would also be of interest. Considering these motivations, we have two main purposes in this paper. First, we discuss a list of criteria for the reliability of the predictions of the classical, non-perturbative $\delta N$ formalism and develop a practical prescription for its usage in the inflationary models, paying particular attention to the subtleties that may arise  concerning large fluctuations. Second, we explore a variety of single-field non-attractor models of inflation that lead to interesting behavior of the PDF in the large fluctuation limit. To be more specific, we show that a truncation of the tail, a power-law tail, a double-exponential tail, and a doubly peaked distribution are all possible to be realized for $\zeta$ in single-field non-attractor models of inflation. We conclude that, generically, the non-perturbative effects significantly alter the tail of the distribution of the primordial fluctuations in most models of inflation that are capable of generating a large abundance of PBHs.\footnote{Note that here we do not study  the canonical, slow-roll, attractor models of inflation since they are generically too simple to generate a significant number of PBHs. Thus, we contend, our generic conclusion is not affected by this exclusion.} As a result, PBH abundance estimations based solely on the perturbation theory may not be reliable, unless it is justified in certain---but perhaps limited---situations.
	
	The rest of the paper is organized as follows.  In Sec.~\ref{sec:method} we review the method of Ref.~\cite{Hooshangi:2021ubn} for obtaining the PDF using the $\delta N$ formalism, starting with the basics in Sec.~\ref{sec:basics}.  We then allude to the underlying assumptions in Sec.~\ref{sec:assumptions} and point out the subtleties and their resolutions in Sec.~\ref{sec:deltaN_large} which we clarify in a working example in Sec.~\ref{sec:example_subtle}.  In Sec.~\ref{sec:quantify-tail} we introduce measures to quantify the behavior of the PDF tail.  Sec.~\ref{sec:models} is devoted to presenting models with non-trivial tail.  This includes studying linear potential (Sec.~\ref{sec:linear}),  quadratic potential (Sec.~\ref{sec:quadratic}), a potential that leads to a power-law tail (Sec.~\ref{sec:power-law}) and a potential barrier (Sec.~\ref{sec:barrier}). We conclude in Sec.~\ref{sec:conclusions}.
	
	Throughout this paper, we set $M_{\rm p}=1/\sqrt{8\pi G}=1$.
	
	\section{A non-perturbative method for computing the full PDF}\label{sec:method} % and notation 
	
	In this section we review the method of non-perturbative calculation of  the probability distribution outlined in Ref.~\cite{Hooshangi:2021ubn}, discuss some subtleties that may arise and argue for some possible resolutions.
	
	\subsection{Basics of the method: small fluctuations}
	\label{sec:basics}
	
	Our method is based on the $\delta N$ formalism which is shown to hold non-perturbatively for super-horizon fluctuations~\cite{Sasaki:1995aw, Wands:2000dp, Lyth:2004gb, Sugiyama:2012tj, Abolhasani:2019cqw}. We restrict ourselves to the single-field inflation (but note that a generalization to the multiple-field is straightforward \cite{Hooshangi:2022lao}). For a given potential $V(\phi)$, the background Klein-Gordon equation in terms of the number of $e$-folds $n$ reads\footnote{We denote the $e$-folds as the time variable by $n$ and the $e$-folds that take for the field to reach some final value from a given initial condition by $N$. The latter, would be the relevant quantity to the $\delta N$ formalism.}
	\begin{equation}\label{KG}
	\phi'' + \left( 3 - \frac12 \phi'^2 \right) \left( \phi' + \frac{V_{,\phi}}{V} \right) = 0,
	\end{equation}
	where prime denotes $d/dn$.\footnote{For brevity, when there is no confusion, we also use prime to denote the derivative of potential with respect to the field.} In an inflationary (but not necessarily attractor) background for which $\phi'^2\ll 1$ we may approximate this exact equation by 
	\begin{equation}
	\label{KG_approx}
	\phi'' + 3   \phi' + 3\frac{V_{,\phi}}{V} \simeq 0.
	\end{equation}
	In this paper, we will use Eq.~\eqref{KG} for our numerical calculations and Eq.~\eqref{KG_approx} for most of our analytic estimates.
	We choose the unperturbed initial field value $\bar\phi$ to correspond to the horizon crossing of the PBH scale of interest during inflation.  We also denote by $\phi=\phi_e$ the end of inflationary era or the beginning of a new phase in which the curvature perturbations are conserved.  Throughout this paper, we assume that $\bar\phi>\phi_e$, so that the net motion of the field is a rolling from large to small values (but we allow for the possibility of a temporary roll to the large values, depending on the field's initial velocity which can have either sign). 
	We also assume that $\phi_e$ is sufficiently far from $\bar \phi$ so that large fluctuations that are already past $\phi_e$, i.e. $\delta\phi < \delta\phi_e \equiv \phi_e-\bar\phi$,  may be realized with extremely low likelihood and can be neglected. Note that this is  indeed a general assumption behind the $\delta N$ formalism. While in the  perturbative  application of the $\delta N$ formalism this assumption is easily satisfied, for our purpose, i.e., studying large fluctuations, this is a non-trivial requirement that needs to be fulfilled.
	
	In the $\delta N$ formalism the quantity of interest is the number of $e$-folds $\N(\bar \phi+\delta \phi)$ between the initial hypersurface $\phi = \bar\phi+\delta\phi$ and the final hypersurface $\phi=\phi_e$.  This is readily found by solving Eq.~\eqref{KG} and then requiring that $\phi(n = \N(\bar \phi+\delta \phi)) = \phi_e$. We hold the initial velocity $\bar\pi \equiv \phi'(0)$ unperturbed since its fluctuations decay exponentially at super-horizon scales \cite{Namjoo:2012aa}.  However, note that we must take into account the dependence of $N$ on $\bar \pi$, since we do not assume an  attractor behavior.  Knowing $\N(\bar \phi+\delta \phi)$, leads to a non-perturbative expression for the curvature perturbation on uniform hypersurfaces which we denote by $\zeta$. We have\footnote{Without subtracting $\langle\delta\N\rangle$, $\zeta$ would be zero over those patches that have an unperturbed history, i.e., $\delta\phi=0$.  But we want $\zeta$ to have vanishing \textit{average} and absorb the residual $\langle\delta\N\rangle$ into the background scale factor.}
	\begin{equation}
	\label{zeta_def}
	\zeta = \delta\N - \langle\delta\N\rangle,
	\end{equation}
	where
	\begin{equation}\label{deltaN}
	\delta\N = \N(\bar\phi + \delta\phi) - \N(\bar\phi).
	\end{equation}
	We can use this to find the PDF of $\delta \N$ or $\zeta$. Assuming a Gaussian PDF of $\delta\phi$, namely
	\begin{equation}\label{PDF-delta-phi}
	\rho_{\delta\phi} = \frac{1}{\sqrt{2\pi} \sigma_{\delta\phi}} e^{-\delta\phi^2/2\sigma_{\delta\phi}^2},
	\end{equation}
	where $\sigma_{\delta\phi}=H/2\pi$ evaluated at the initial time (i.e., when the mode of interest left the horizon) we have
	\begin{equation}\label{PDF-zeta-vs-delta-phi}
	\rho_{_{\delta \N}} = \left| \frac{d\delta\phi}{d\delta \N} \right| \rho_{\delta\phi} = \frac{1}{|\N'(\bar\phi + \delta\phi)|} \rho_{\delta\phi},
	\end{equation}
	which reflects the non-linear relationship between $\delta\phi$ and $\delta \N$ in the non-Gaussian shape of $\rho_{_{\delta \N}}$.

	\subsection{Underlying assumptions}
	\label{sec:assumptions}
	
	In this section, we briefly discuss the underlying assumptions that have to be fulfilled for the $\delta \N$ formalism, as discussed in the previous section, to work. These requirements are also outlined in \cite{Hooshangi:2021ubn}.
	
	The first assumption is the Gaussianity of $\rho_{\delta\phi}$.  This can be ensured by smallness of ${\cal L}_n/{\cal L}_2$, where ${\cal L}_n$ is the $n$th term in the Lagrangian of $\delta\phi$. For ${\cal L}_2$ we have
	\begin{equation}
	{\cal L}_2 = \frac12 \delta\dot\phi^2-\frac{1}{2a^2} (\nabla \delta\phi)^2  + \frac12 V''(\bar\phi) \delta\phi^2 \sim \frac12 H^2 \delta\phi^2 \approx \frac16 V(\bar\phi) \delta\phi^2,
	\end{equation}
	where we have estimated $\delta\dot\phi$ and $\nabla \delta \phi/a$ by $H\delta\phi$ (since we are interested in computing ${\cal L}_2$ around the horizon crossing time), noted that the mass term is at most of the same order as the other two terms, and assumed that either the first or the second term represents the size of ${\cal L}_2$.
	
	Regarding the interaction Lagrangian, often times it is the matter sector, rather than the gravity sector, that dominates.
	For an analytic potential which can be Taylor expanded around $\bar\phi$, the third-order Lagrangian is ${\cal L}_3  \sim \frac{1}{6} V'''(\bar\phi) \delta\phi^3$.  Therefore, the ratio of interest is 
	\begin{equation}\label{L3/L2}
	\frac{{\cal L}_3}{{\cal L}_2} \sim \frac{V'''(\bar\phi)}{V(\bar\phi)} \delta\phi.
	\end{equation}
	
	When the potential does not have a Taylor expansion up to the third order (like when $V\propto(\phi-\bar \phi)^{ q}$ with $2<q<3$, which shall be studied in Sec.~\ref{sec:power-law}), instead of $V'''(\bar\phi)\delta\phi^3$, we use 
	\begin{equation}
	\label{L3/L2*}
	\frac{V(\bar\phi+\delta\phi) -  \left[ V(\bar\phi) + V'(\bar \phi)\delta \phi + \frac12 V''(\bar \phi) \delta \phi^2 \right]}{V(\bar\phi) \delta\phi^2}\, ,
	\end{equation}	
	where the numerator is the full potential with 
	the up-to-quadratic terms subtracted.	If $V'(\bar\phi)=V''(\bar\phi)=0$ (which would be the case for $V\propto(\phi-\bar \phi)^{ q}$ with $2<q<3$), this simplifies to
	\begin{equation}
	\label{L3/L2_nonanalytic}
	\frac{V(\bar\phi+\delta\phi) - V(\bar\phi)}{V(\bar\phi) \delta\phi^2}.
	\end{equation}
	We will check the smallness of either~\eqref{L3/L2} or \eqref{L3/L2*} for each of our examples in the subsequent sections.  Finally, note that, unlike the Lagrangian of $\delta\phi$, the Lagrangian of $\zeta$ shows non-perturbative behavior; so its higher order terms are not necessarily small, and it is the non-perturbative nature of the $\delta N$ formalism that captures these effects.
	
	The second assumption is the smallness of stochastic effects, whose relative importance over a classical field excursion $\Delta\phi$ and during $\N$ $e$-folds, is controlled by 
	\begin{equation}
	\label{S}
	{\cal S} = \frac{H}{2\pi\Delta\phi} \sqrt{\N}.
	\end{equation}
	To see this, just note that during this excursion, the expected displacement due to the accumulated jumps is given by the step size ($H/2\pi$) multiplied by the random walk factor $\sqrt{\N}$.  The smallness of ${\cal S}$ will be checked in the  examples that follow. However, in certain situations  this criterion to estimate the significance of stochasticity may not be applicable, in  which case we present a different argument.
	
	Notice that both ${\cal L}_3/{\cal L}_2$ and ${\cal S}$ depend on the size of fluctuations that one considers so that they eventually become large for sufficiently large fluctuations (if such large fluctuations are allowed). In the examples that follow, we present the values of these quantities for the values of $\delta \phi$ corresponding to $\zeta \sim 1$, which is  roughly the region most relevant to the PBH formation.
	
	The third assumption is neglecting the fluctuation $\delta\pi$ of the momentum, which is justified since the mode function of the massless field $\delta\phi$ is frozen at super-horizon scales.
	
	Provided that the above assumptions hold, the  procedure outlined in Sec.~\ref{sec:basics} is  sufficient to handle most situations, but there are refinements that are in principle important when large fluctuations are considered and which we explain below.

	\subsection{$\delta N$ for large fluctuations}
	\label{sec:deltaN_large}
	
	In this section, we discuss the subtleties that may arise  when one studies the statistics of large fluctuations using the $\delta \N$  formalism. This leads to some additional ingredients one needs to add to the simple procedure outline in Sec.~\ref{sec:basics}. 
	
	In the first step, one has to inspect the potential to see if there exists an upper bound $\delta\phi_\text{max}$ on the allowed fluctuations.  This may be due to circumstances that prohibit reheating (or exhibit eternal inflation)---e.g., due to not being able to reach $\phi_e$---or any other reason that may invalidate the theory beyond $\delta\phi_\text{max}$.   Therefore, our analysis will be applicable to a fraction
	\begin{equation}
	\label{pr}
	p_r \equiv  \int_{-\infty}^{\delta\phi_\text{max}} \rho_{\delta\phi} \, d\delta\phi  = \dfrac12 \left[ 1 + \text{erf} \left( \frac{\delta \phi_{\max}}{\sqrt{2}\sigma_{\delta \phi}} 
	\right) \right]
	\end{equation}
	of the field fluctuations and so we have to condition our probabilities on $ \delta\phi < \delta\phi_\text{max}$, or equivalently, work with $\frac{1}{p_r} \rho_{\delta\phi}$ instead of $\rho_{\delta\phi}$ itself.  Strictly speaking, the lower limit is not $-\infty$ either since our formalism cannot describe the fluctuations that are so large and negative that are past $\phi_e$ without any classical evolution. However, since $\phi_e$ is arbitrary, we can always choose it to be far enough from $\bar \phi$  so that the lower bound becomes irrelevant. Note that the upper bound may also become irrelevant if $\delta\phi_\text{max} \gg \sigma_{\delta \phi}$ but need not be the case in specific models. Depending on the details of the model, there are two possibilities regarding large and positive values of $\delta \N$. If $\delta \N(\delta\phi_\text{max}) \equiv \delta\N_\text{max}<\infty$, there would be a truncation in the PDF and $\delta \N$ larger than $\delta\N_\text{max}$ cannot be realized. If $ \delta\N_\text{max}=\infty$, there would be no truncation.\footnote{It is interesting to notice that the absence of upper bound on the field fluctuations (i.e., if $\delta\phi_\text{max} \to \infty$) does not imply the absence of truncation in $\rho_{_{\delta \N}}$. That is, we might have $\delta \N_{\max}=\delta N(\delta\phi \to \infty)< \infty$. We present an example of this situation in Sec.~\ref{sec:quadratic}.} 
	
	As the next step, we may simplify the situation by noticing that studying the model from $\bar \phi$ all the way down to $\phi_e$---which requires the knowledge of the potential over this large range---is often unnecessary. In particular, we consider a non-attractor phase of inflation from $\bar \phi$ to $\phi_c$ (satisfying $ \phi_e < \phi_c <\bar \phi$)  and assume an abrupt transition to a slow-roll phase for $\phi<\phi_c$. Since in the slow-roll phase, $\zeta$ will be frozen, it suffices to keep evolving the fluctuations until $\phi_c$ is reached. To ensure that the evolution terminates immediately after $\phi_c$, we require to have $|\bar \pi| \ll \sqrt{2\epsilon_{v}}$  where $\epsilon_{v}\equiv V'^2/2V^2$ is the first slow-roll parameter in the slow-roll phase. The argument for this claim is based on the intuition from the $\delta \N$ formalism and is the following. Since  $|\bar \pi|$ is small compared to the attractor field velocity $\sqrt{2\epsilon_{v}}$ and the non-attractor evolution reduces the velocity even further, all trajectories that start from $\phi>\phi_c$ enter the slow-roll phase with almost zero velocity (compared to the attractor velocity). As a result, the number of $e$-folds from $\phi_c$ to $\phi_e$ becomes independent of the initial field value and this range can be ignored in computing $\delta \N$.\footnote{Note that the condition $|\bar \pi| \ll \sqrt{2\epsilon_{v}}$  is consistent with the ``sharp transition'' criteria discussed in Ref.~\cite{Cai:2018dkf} that ensures that the transition to the slow-roll phase does not spoil the non-trivial statistics achieved in the non-attractor phase.}
	
	The above simplification, i.e., counting the number of $e$-folds up to $\phi_c$ rather than $\phi_e$ may result in a subtlety that needs to be dealt with. Define $\delta \phi_e \equiv \phi_e-\bar \phi$ and $\delta \phi_c \equiv \phi_c - \bar \phi$. While we have chosen $\phi_e$ so that $\delta \phi_e \gg \sigma_{\delta \phi}$ it is not necessarily the case that $\delta \phi_c \gg \sigma_{\delta \phi}$. This implies that there might be perturbed trajectories that would be of our interest but which start from $\phi<\phi_c$ (corresponding to the large and negative fluctuations, $\delta \phi < \delta \phi_c$). To non-perturbatively describe such fluctuations one needs to know the details of the potential for $\phi<\phi_c$ which we tried to avoid! However, since $\phi<\phi_c$ is the slow-roll regime, such trajectories totally miss the non-attractor phase. Thus, we expect the linear perturbation theory to remain reliable for a rather broad range of  fluctuations satisfying $\phi \lesssim \delta \phi_c$. This leads us to approximate the relation between $\delta \N$ and $\delta \phi$ by a Taylor expansion up to linear order around $\delta \phi_c$ (without assuming $\delta \phi$ to be small). We have 
	\begin{equation}\label{N-around-phic}
	\delta \N(\delta \phi) \approx \delta \N_c + \dfrac{\sigma_c}{\sigma_{\delta \phi}} (\delta\phi - \delta\phi_c)\, ,\quad \text{for $\delta\phi \lesssim \delta \phi_c$,}
	\end{equation}
	where $\delta \N_c$ and $\sigma_c$ are two parameters to be determined. This linear relation implies that for $\delta \phi <\delta \phi_c$ (or, equivalently, for $\delta \N < \delta \N_c$)  we have approximated the PDF of $\delta \N$ by a Gaussian distribution centered at $\delta \N_c -\sigma_c \delta \phi_c/\sigma_{\delta \phi}$ and with width $\sigma_c$.   We thus have the following full PDF for $\delta \N$:
	\begin{equation}
	\label{full_pdf}
	\rho_{_{\delta \N}} =
	\begin{cases} 
	\displaystyle
	\rho_{_{\delta \N}}^+ = \frac{1}{p_r} \frac{1}{|\N'(\bar\phi + \delta\phi)|} \rho_{\delta\phi} \hspace{5.37cm}  \text{ $\delta\N > \delta\N_c$}\, ,
	\vspace{.5cm}
	\\
	\displaystyle
	\rho_{_{\delta \N}}^- = \frac{1}{p_r} \frac{1}{\sqrt{2\pi}\sigma_{c}} \exp \left\{ -\frac{1}{2\sigma_{c}^2} \left[ \delta\N - \delta\N_c + \sigma_c \delta\phi_c/\sigma_{\delta \phi} \right]^2 \right\}\, \quad \text{ $\delta\N < \delta\N_c$}\, ,
	\end{cases}
	\end{equation}
	where $p_r$ is defined in Eq.~\eqref{pr}.
	$\rho_{_{\delta \N}}^+ $ is the non-perturbative PDF appropriate for fluctuations satisfying $\delta\N > \delta\N_c$ (or $\delta \phi > \delta \phi_c$) for which it suffices to count the $e$-folds until $\phi_c$ is reached (assuming $|\bar \pi| \ll \sqrt{2\epsilon_{v}}$, see the discussion of the previous paragraph). On the other hand, $\rho_{_{\delta \N}}^-$ is a Gaussian distribution, approximating the PDF for fluctuations that do not experience the non-attractor phase, i.e., the ones with $\delta\N < \delta\N_c$ (or $\delta \phi < \delta \phi_c$). 
	
	To fully determine the above PDF, we need to determine two unknown parameters, $\delta \N_c$ and $\sigma_c$. The continuity of $\delta \N$, results in $\delta \N_c =\N(\phi_c)-\N(\bar \phi)$ where, as discussed,  $\N$ on the right hand side can be computed up to $\phi_c$. Thus the first term vanishes and the second term is the $e$-folds from $\bar \phi$ to $\phi_c$, which we denote by $\bar N_c$.
	
\begin{figure}
		\center 
		\hspace{-2cm}
		\includegraphics[scale=.5]{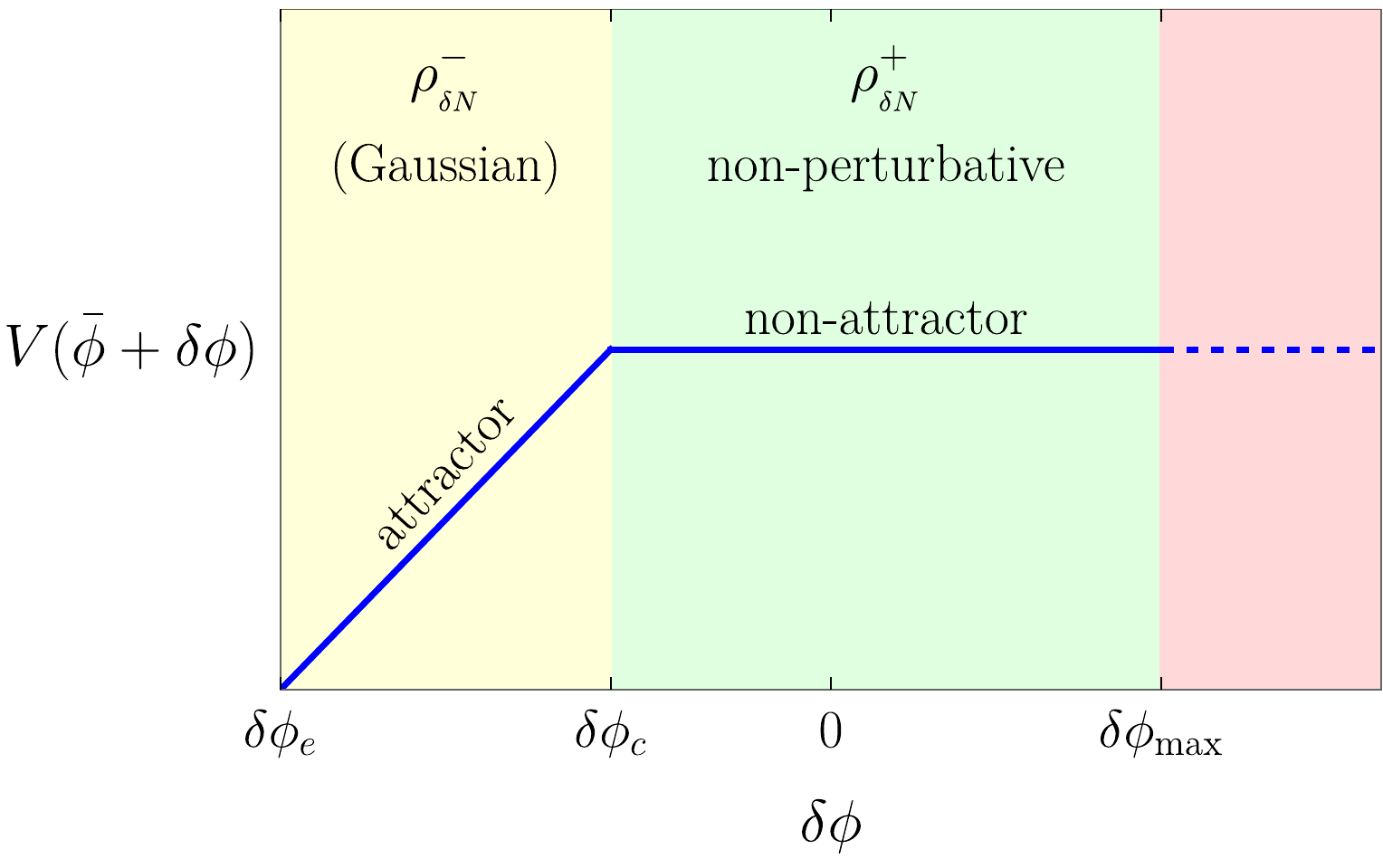}
		\caption{This plot summarizes our assumptions and the way we treated large fluctuations, leading to the PDF Eq.~\eqref{full_pdf}. The blue curve depicts the potential in different phases while different shaded regions correspond to different treatments of fluctuations. The red region corresponds to the fluctuations larger than $\delta \phi_{\max}$ so that they must be excluded (since, e.g., the trajectories start from this region cannot reach the end point $\bar \phi +\delta \phi_e$). In the green region, a non-attractor evolution takes place. The fluctuations in the green region are treated non-perturbatively leading to a non-perturbative PDF for $\delta \N$, denoted by $\rho_{_{\delta \N}}^+$. For these fluctuations it suffices to count $e$-folds until $\phi_c$ is reached (since the $e$-folds from $\phi_c$ to $\phi_e$ for trajectories emanating from the green region are almost the same and do not contribute to $\delta \N$). In the yellow region we assume a slow-roll potential with sufficiently large slope (corresponding to sufficiently large attractor velocity as compared with the initial velocity $\bar \pi$). For the trajectories that start from the yellow region  slow-roll conditions hold in the entire history so that the linear relation between $\delta \phi$ and $\delta \N$ is justified. We thus Taylor expand this relation for fluctuations close to (but smaller than) $\delta \phi_c$  and keep up to the linear term. This leads to the $\rho_{_{\delta \N}}^-$ piece of the PDF.}
		\label{fig:rho_sum}
	\end{figure}

	Next, notice that since the trajectories corresponding to $\delta\N < \delta\N_c$ only experience the slow-roll phase, we expect the width of the PDF for these fluctuation to be determined by the slope of the potential, i.e., $\sigma_c\simeq \sigma_{\delta \phi}/\sqrt{2\epsilon_{v}}$. Thus, we have fixed the two remaining unknown parameters to be
	\begin{equation}
	\delta \N_c \simeq  -\bar N_c
	%-\bar N(\phi_c)
	\, , \qquad
	\sigma_c  \simeq \dfrac{\sigma_{\delta \phi}}{\sqrt{2\epsilon_{v}}}.
	\end{equation}
	We conclude that, for all practical purposes, the only information one needs about the slow-roll phase is the slope of the potential, encoded in $\epsilon_{v}$. Throughout this paper, we choose $\sqrt{2\epsilon_{v}}=10 |\bar \pi|$. Thus, for the examples that we shall discuss, it suffices to only report $\bar \pi$. Figure.~\ref{fig:rho_sum}  summarizes the assumptions and arguments that led to the PDF Eq.~\eqref{full_pdf}.
	
	Note that despite the fact that we excluded the realization of the inflaton's fluctuations beyond $\delta \phi_{\max}$, the full PDF for $\delta \N$ (Eq.~\eqref{full_pdf}) is, by construction, normalized to one. In many situations we have $|\delta \phi_c| \gtrsim \sigma_{\delta \phi}$ (which is equivalent to sufficiently large $\bar N_c$) in which case  $\rho_{_{\delta \N}}^-$ would be irrelevant for the practical purposes. However,  we will see that in certain situations this condition does not hold and the Gaussian piece plays an important role e.g., in the PDF being normalized. Further notice  that $\rho_{_{\delta \N}}$ is continuous at $\delta \N_c$ only if $|\N'(\phi_c)| =1/\sqrt{2\epsilon_{v}}$ which  does not need to be satisfied. Thus, generically, we expect to have a discontinuity in the PDF which is a  consequence of the break in the potential at $\phi_c$. 
	
	After obtaining the PDF for $\delta \N$ our last step is to compute the PDF for $\zeta$ by a shift according to Eq.~\eqref{zeta_def}. That is, we have
	\begin{equation}
	\zeta = \delta\N - \langle\delta\N\rangle \qquad \Longrightarrow \qquad \rho_\zeta(\zeta) = \rho_{_{\delta \N}}(\langle\delta\N\rangle + \zeta).
	\end{equation}
	
	This result generically leads to a non-trivial right tail of PDF for $\zeta$  while the left tail remains Gaussian. This is a result of our assumed setup in which we start from a non-attractor phase and end with an attractor phase. Note that this is appropriate for the PBH formation since overdensities would be sensitive to the right tail of the distribution.

	We now examine how well the above approximated PDF works in an explicit example.
	
	\subsubsection{A working example}
	\label{sec:example_subtle}
	
\begin{figure}
		\center
		\includegraphics[scale=.55]{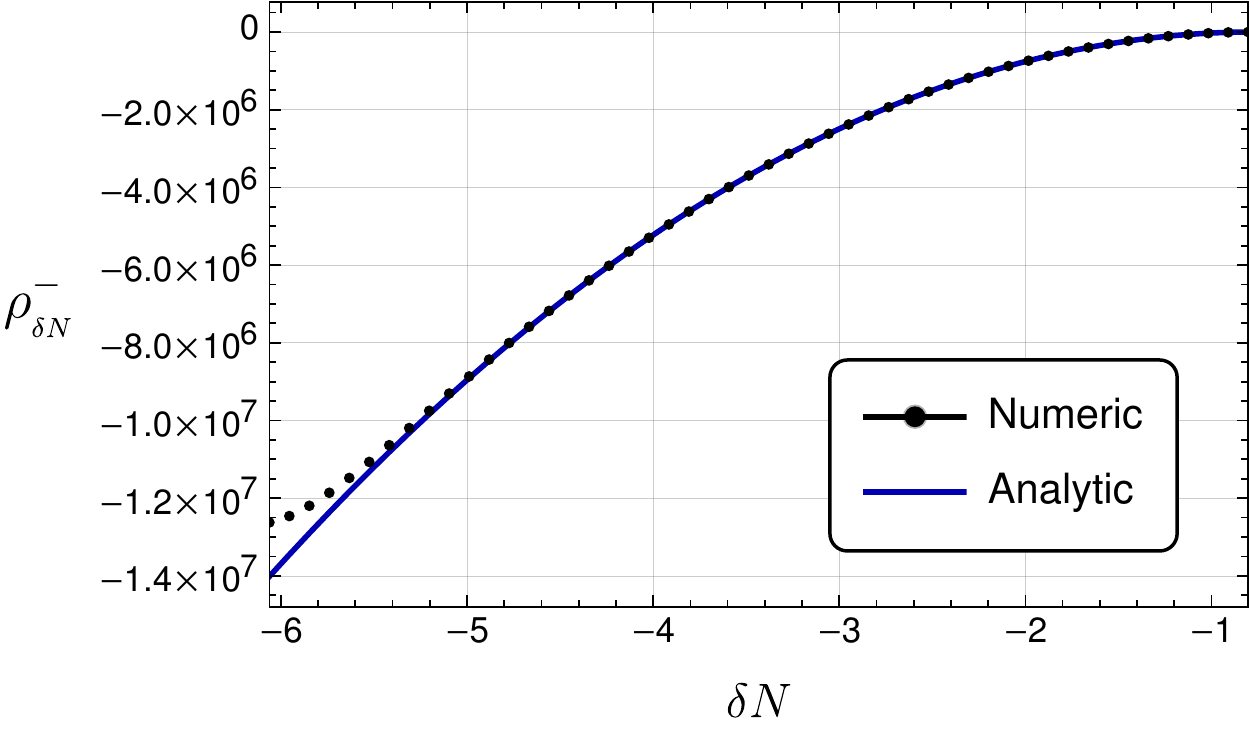}
		\hspace{.5cm}
		\includegraphics[scale=.509]{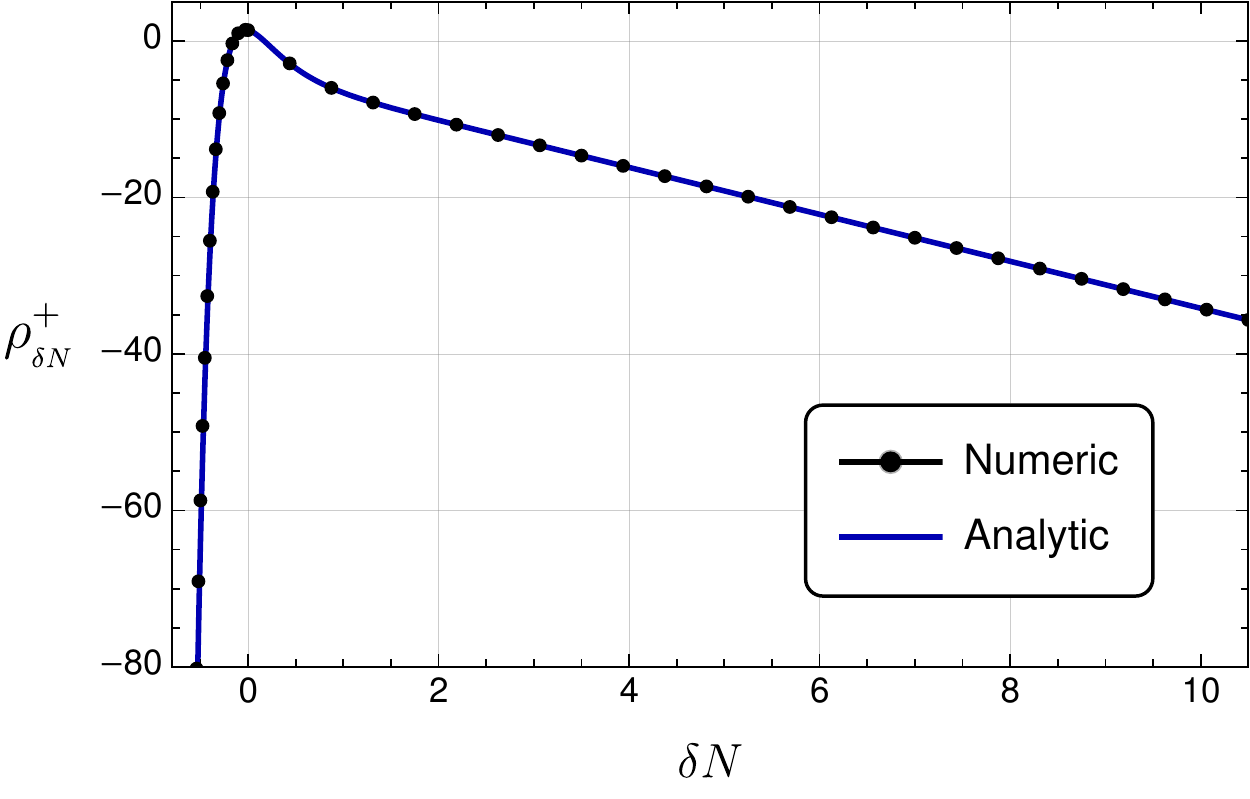} 
		\caption{Comparison between the numerical result and the analytic approximation (Eq.~\eqref{full_pdf_example}) for the two pieces of PDF. We have set $\bar{\phi}=3 \times 10^{-4}$, $\phi_c = -5 \times 10^{-2}$, $\bar\pi=-10^{-3}$, $\alpha=10^{-2}$ and $V_0=1.18 \times 10^{-8}$. This choice of parameters correspond to $\delta \N_c\simeq -\bar N_c\simeq -0.77$. }
		\label{fig:rho_check}
	\end{figure}

	 In this section, we study a simple example to compare our approximated PDF Eq.~\eqref{full_pdf} with the exact results which we obtain numerically. What we consider here is basically the well-studied ultra slow-roll (USR) model with the additional slow-roll  phase. Suppose that the potential is given by
	\begin{equation}
	\label{V_subtle}
	V(\phi) = \begin{cases}
	V_0, & \phi>\phi_c, \\
	V_0 \left[ 1 + \alpha (\phi-\phi_c) \right], & \phi\leq \phi_c\, ,
	\end{cases}
	\end{equation}
	where we only consider $\alpha>0$ here and assume $| \alpha (\phi-\phi_c) |\ll 1$ for the entire evolution and for all trajectories that we consider. Further
	assume that the unperturbed trajectory is such that the inflaton starts from $\bar\phi>\phi_c$ with negative initial velocity $\bar\pi<0$, experiences an USR phase, and then quickly falls in the attractor slow-roll regime after passing $\phi_c$ until reaching $\phi_e<\phi_c$.  Under the above assumptions one may solve Eq.~\eqref{KG_approx} for the potential Eq.~\eqref{V_subtle} which results in
	\begin{equation}\label{USR-phi-N}
	\phi(n) \simeq \begin{cases} 
	\phi_0 + \frac13 \bar \pi \left(1 - e^{-3 n} \right), & \phi>\phi_c \, \, \& \, \,\phi_0 >\phi_c  \\ 
	\phi_c+ \frac13 (\pi_c+\alpha) \left( 1-e^{-3(n-n_c)} \right) - \alpha (n-n_c), & \phi\leq \phi_c  \, \, \& \, \,\phi_0 >\phi_c \\
	\phi_0+ \frac13 (\bar \pi+\alpha) \left( 1-e^{-3n} \right) - \alpha n, &    \phi_0 \leq  \phi_c\, ,
	\end{cases}
	\end{equation}
	where $\phi_0 \equiv \bar \phi +\delta \phi$ and $\bar \pi$ are the initial conditions  and we have defined 
	\begin{equation}\label{N_c}
	n_c \equiv  -\frac13 \log \left[ 1 + 3\frac{\phi_0-\phi_c}{\bar\pi} \right]  \, , \quad \pi_c \equiv  \bar\pi e^{-3n_c} = \bar \pi +3(\phi_0-\phi_c).
	\end{equation}
	$n_c$ is the number of $e$-folds that takes for the inflaton to reach $\phi_c$ from $\phi_0$ and $\pi_c$ is the field velocity at $\phi=\phi_c$. Using $\phi(\N)=\phi_e$, one can obtain an implicit relation for $\N(\bar \phi+\delta \phi)$ for all values of $\delta \phi$ from which one can also compute the full PDF. However, we are interested in studying the validity of Eq.~\eqref{full_pdf} which, for our specific example, reduces to
	\begin{equation}
	\label{full_pdf_example}
	\rho_{_{\delta \N}} \simeq 
	\begin{cases} 
	\displaystyle
	\rho_{_{\delta \N}}^+ = \frac{1}{p_r}  \frac{|\bar\pi_c|}{\sqrt{2\pi}\sigma_{\delta\phi}} e^{-3  \delta\N} \exp \left\{ -\frac{\bar \pi_c^2}{18\sigma_{\delta\phi}^2}  \left( e^{-3\delta\N} - 1 \right)^2 \right\}, \, \qquad \text{$\delta\N >- \bar N_c$}
	\vspace{.4cm}
	\\
	\displaystyle
	\rho_{_{\delta \N}}^- = \frac{1}{p_r} \frac{\alpha}{\sqrt{2\pi}\sigma_{\delta\phi}} \exp \left\{ -\frac{\alpha^2}{2\sigma_{\delta\phi}^2} \left[ \delta\N +\bar N_c + \alpha^{-1} \delta\phi_c \right]^2 \right\} , \qquad \text{$\delta\N <-\bar N_c$}.
	\end{cases}
	\end{equation}
	where ${\bar \pi}_c=\bar \pi +3(\bar \phi -\phi_c)$ is the field  velocity at $\phi_c$ for the unperturbed trajectory and $\bar N_c= N(\bar \phi)$. The quantity $p_r$ is defined via Eq.~\eqref{pr} with $\delta\phi_\text{max} \equiv \phi_c - \frac13\bar\pi - \bar\phi$. If $\delta \phi > \delta\phi_\text{max}$ it is classically impossible for the field to reach $\phi_e$ and eternal inflation occurs.  Such large fluctuations are therefore to be excluded since our analysis of the PBHs is implicitly conditioned on the assumption that reheating does take place. Notice that for large and positive $\delta \N$ the PDF behaves as $e^{-3\delta \N}$ which is the tail behavior for the USR models of inflation \cite{Biagetti:2021eep}. In Figure~\ref{fig:rho_check} we compare $\rho_{_{\delta \N}}^-$ and $\rho_{_{\delta \N}}^+$ of Eq.~\eqref{full_pdf_example} with the full numerical results which demonstrates an excellent agreement for a broad range of fluctuations.

	\subsection{Quantifying the tail behavior of the PDF}
	\label{sec:quantify-tail}
	
	To investigate the tail of the PDF, we introduce two quantities that we may compute for different examples that follow. The first quantity is the probability that fluctuations larger than some critical value can be realized. Working with $\zeta$ and choosing unity as the critical value, we define
	\begin{equation}
	\label{beta_def}
	\beta_\zeta \equiv \int_{1}^{\infty} \rho_{\zeta}\, d\zeta\, .
	\end{equation}
	Notice that this is not directly related to the PBH abundance for which one needs to work with the density contrast rather than the curvature perturbation. Since this step is already well established and can be found e.g., in Refs.~\cite{Musco:2018rwt,Biagetti:2021eep} we do not discuss it in this paper. However, $\beta_\zeta$, as defined according to Eq.~\eqref{beta_def}, is a simple tool to estimate how wildly the tail behaves as compared, e.g., with a fully Gaussian PDF as predicted by the linear perturbation theory.
	
	The second quantity that we introduce to further explore the tail is
	\begin{equation}
	\label{D}
	{\cal D} = -\frac{d\log\rho_\zeta}{d\zeta} = -\frac{d\log\rho_{_{\delta \N}}}{d\delta \N}.
	\end{equation}
	$\cal D$ behaves differently for large $\zeta$ for different tails which can be seen in a few simple examples. Define ${\cal D}_\infty \equiv \lim_{\zeta\to+\infty} {\cal D}$ then: (i) A Gaussian PDF $\propto e^{-\zeta^2/2\sigma^2}$ has ${\cal D}_\infty  = \infty$. (ii) An exponential PDF $\propto e^{-k\zeta}$ has ${\cal D}_\infty = k > 0$. (iii) A power-law PDF $\propto 1/\zeta^p$ has ${\cal D}_\infty  = 0$. (iv) A PDF $\propto e^{-k\zeta^p}$ with $1>p>0$ has ${\cal D}_\infty  = 0$. What we dubbed a practically heavy-tailed PDF in \cite{Hooshangi:2021ubn} has the property that ${\cal D}_\infty = 0$.
	Thus our definition accepts examples (iii) and (iv) as practically heavey-tailed, but excludes examples (i) and (ii).

	In the rest of this paper we apply the method outlined in Sec.~\ref{sec:deltaN_large} and confirmed in Sec.~\ref{sec:example_subtle} to various examples and see that non-trivial tails are generic. 
	
	\section{Non-attractor models of inflation exhibiting non-trivial tail}
	\label{sec:models}
	
	In this section, we try to establish what we advocated: We show that the tail behavior can vary significantly when a non-attractor phase occurs in single-field models of inflation. Here we focus on the non-perturbative evaluation of  $\rho_{_{\delta \N}}^+$ since the other piece $\rho_{_{\delta \N}}^-$ is a simple Gaussian PDF. Thus, from now on, $\N(\phi)$ will be measured from $\phi$ to $\phi_c$, not to $\phi_e$. While the $\rho_{_{\delta \N}}^-$ piece will not be shown in the graphs, it will be taken into account for  the correctness of the normalization which---in some examples---also affects the values of different quantities that we defined in Secs.~\ref{sec:assumptions} and~\ref{sec:quantify-tail} and we compute in the following examples. Finally, since $\zeta$ differs from $\delta \N$ only by a shift, we mainly present the results in terms of $\delta \N$ as it is more directly related to the background solutions and easier to obtain intuition from. However, when we report some values for our parameters such as $\beta_\zeta$ we do make the appropriate shift.
	
	\subsection{General up-to-quadratic potential}
	\label{sec:general-quadratic}
	
	We start by a general result from an analytic potential which may be Taylor expanded.  In this section, we consider up to the quadratic term in the expansion and study the special cases in the two subsequent sections. The potential may be quantified by
	
	\begin{equation}\label{general-V2}
	V = V_0 \left( 1 + \alpha\phi + \frac12 \beta\phi^2 \right).
	\end{equation}
	In what follows, we assume $ |\alpha\phi|\ll 1$ and $ |\beta\phi^2|\ll 1$, but not necessarily $|\beta\phi^2| \ll |\alpha\phi| $. We also assume that this form of potential holds true for $\phi>\phi_c$ (so that higher order terms in the Taylor expansion never become important). 
	The approximate Klein-Gordon equation in a quasi de Sitter universe Eq.~\eqref{KG_approx} for this potential reads
	\begin{equation}\label{KG-quadratic}
	\phi'' + 3\phi' + 3\alpha + 3\beta\phi \simeq 0\, ,
	\end{equation}
	where, we have neglected higher order contributions in $\alpha$ and $\beta$. The solution is given by
	\begin{equation}\label{phi-N-general}
	\phi(n) \simeq  e^{-\frac{3n}{2}} \left\{ \left[ \frac23 \bar \pi + \phi_0 +\phi_a \right] \frac1\Gamma  \sinh \frac{3\Gamma n}{2} + \left( \phi_0 +\phi_a \right) \cosh \frac{3\Gamma n}{2} \right\} - \phi_a,
	\end{equation}
	where, as before,  $\phi_0=\bar\phi+\delta\phi$ and $\bar\pi$ are the initial conditions and we have defined
	\begin{equation}
	\phi_a \equiv \frac{\alpha}{\beta}, \qquad \text{and} \qquad \Gamma \equiv  \sqrt{1-\frac{4\beta}{3}}.
	\end{equation}
	Setting $\phi(\N)=\phi_c$, we obtain
	\begin{equation}\label{delta-phi-zeta-general}
	\delta\phi = \frac{\displaystyle \Gamma e^{\frac{3\N}{2}} \left(\phi_a+\phi_c\right) - \frac23 \bar\pi \sinh \frac{3\Gamma\N}{2}}{\displaystyle \Gamma \cosh \frac{3\Gamma\N}{2} + \sinh \frac{3\Gamma\N}{2} } - \phi_a -\bar\phi,
	\end{equation}
	where $\N$ is to be understood as $\N(\bar \phi+\delta\phi) = \N(\bar\phi) + \langle\delta\N\rangle + \zeta$.  Therefore, this is an implicit relation between $\delta\phi$ and $\zeta$ which, in principle, can be used to obtain the PDF $\rho_\zeta^+$ from $\rho_{\delta\phi}$. Note that if $\beta >3/4$, the same relation holds but with the replacement $\Gamma \to i \tilde \Gamma$ where $\tilde \Gamma \equiv \sqrt{\frac{4\beta}{3}-1}$. This apparently formidable expression is simplified in the special cases of the next two sections. 
	
	\subsubsection{Linear potential}
	\label{sec:linear}

	In this section we consider the very simple case of the linear potential
	\begin{equation}\label{linear-V}
	V = V_0 (1+\alpha\phi).
	\end{equation}
	We shall consider both cases of positive and negative $\alpha$, but in either case, we assume that $\bar \phi>\phi_c$.
	The solution of the equation of motion can be obtained by carefully taking the limit $\beta\to0$ of Eq.~\eqref{phi-N-general}, or by directly solving Eq.~\eqref{KG-quadratic} with $\beta=0$:
	\begin{equation}\label{phi(N)-linear}
	\phi(n) = \phi_0 + \frac13 (\bar \pi+\alpha) \left( 1-e^{-3n} \right) - \alpha n.
	\end{equation}
	
	It follows that
	\begin{equation}
	\label{dphi_N_lin}
	\delta\phi = \frac13 (\bar\pi+\alpha) \left( e^{-3\delta\N}-1 \right) e^{-3\bar \N_c} + \alpha\delta\N,
	\end{equation}
	where $\delta\N = \langle\delta\N\rangle + \zeta$ and, as before, $\bar \N_c =\N(\bar\phi)$ is the number of $e$-folds in the unperturbed evolution from $\bar\phi$ to $\phi_c$, i.e.,
	\begin{equation}\label{phi_c-linear}
	\phi_c = \bar\phi + \frac13 (\bar\pi+\alpha) \left( 1-e^{-3\bar \N_c} \right) - \alpha \bar \N_c.
	\end{equation}
	It is interesting to note that, according to Eq.~\eqref{dphi_N_lin}, $\delta \N$ and $\delta \phi$ have a linear relationship for both  small and  large values of $\delta \N$.\footnote{This statement is valid as long as the limit $\delta\N\to\infty$ exists.  As we will see shortly, in some cases $\delta\N$ is not allowed to take on arbitrarily large values so this limit may not exist.} This implies that the PDF of $\zeta$ is Gaussian both around the peak and far into the tail.  However, this does not mean that a simple extrapolation of the linear theory leads to the correct PDF at its tail. Indeed, the width of the Gaussian behavior around the peak is $\sigma_{\delta \phi}/ \left( \alpha - (\bar \pi +\alpha)e^{-3 \bar \N_c} \right)$ while that at the tail is $\sigma_{\delta\phi}/|\alpha|$. Furthermore, there would be an inevitable  transient behavior between the two Gaussian limits.
	\begin{figure}
		\begin{center}
			\includegraphics[scale=.59]{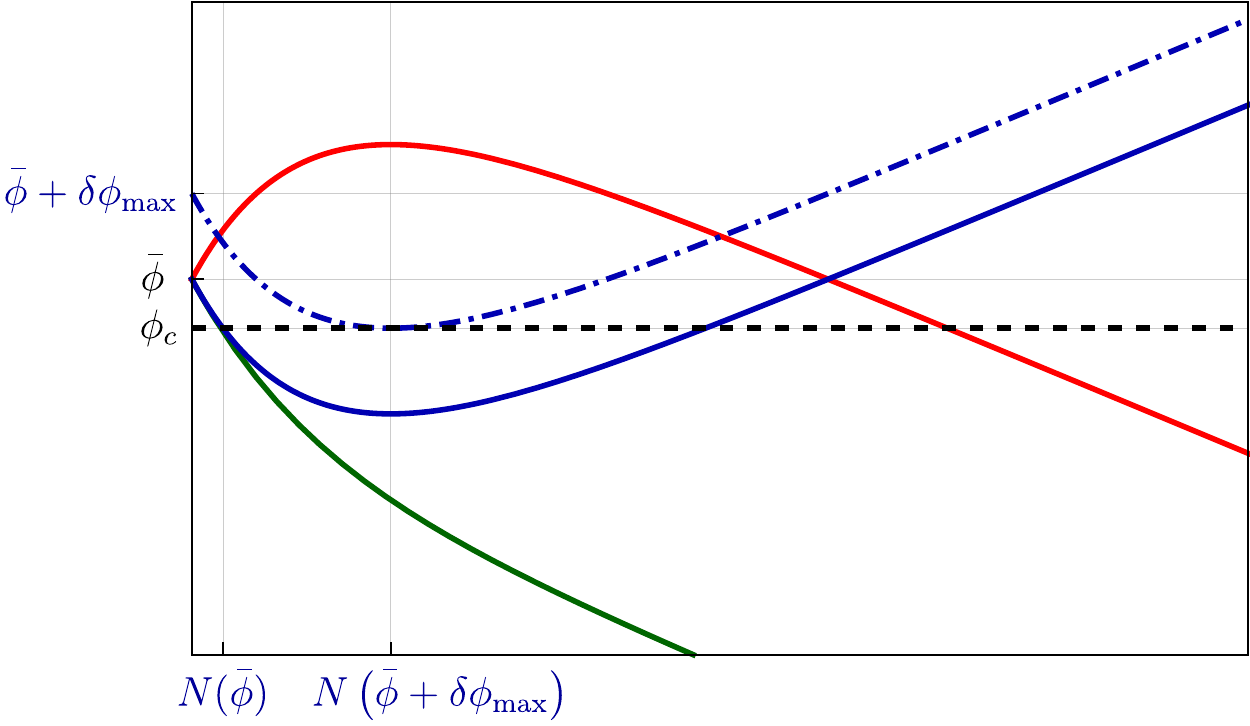}
			\hspace{.25cm}
			\includegraphics[scale=.59]{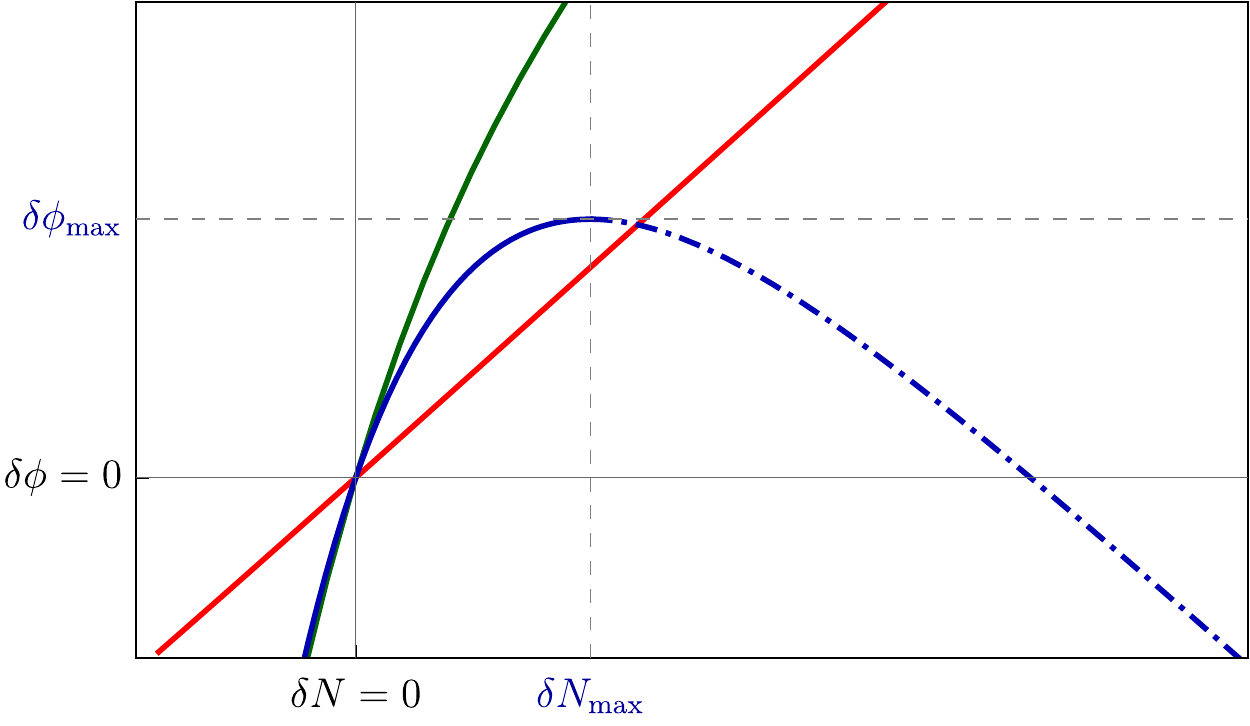}
		\end{center}
		\caption{Left: The schematic time evolution of $\phi(n)$ under the linear potential \eqref{linear-V} for the three cases (i) $\alpha>0$, $\bar\pi<0$, (ii) $\alpha>0$, $\bar\pi>0$ and (iii) $\alpha<0$, $\bar\pi<0$ (green, red and blue, respectively), all with the same $|\alpha|$ and $|\bar\pi|$.  The dashed blue trajectory has the same parameters as the solid blue one, except for $\phi_0=\bar\phi+\delta\phi_{\rm max}$ (instead of $\phi_0=\bar \phi$) which is such that the return point coincides with $\phi_c$.  Each trajectory has a first crossing of $\phi_c$ at an $\N(\phi_0)$. The ticks that are only valid for the case (iii) (i.e., for the blue curves) are  shown in blue. Right: $\delta\phi$ versus $\delta\N$ . Different colors correspond to the three different  cases similar to the left panel and, again, the blue ticks are only valid for the case (iii). The dotted segment is beyond the one-to-one domain of $\delta\phi$-vs-$ \delta\N$ and must therefore be removed.   }
		\label{fig:linear-V-deltaN}
	\end{figure}
	
	\begin{figure}
		\begin{center}
			\includegraphics[scale=.7]{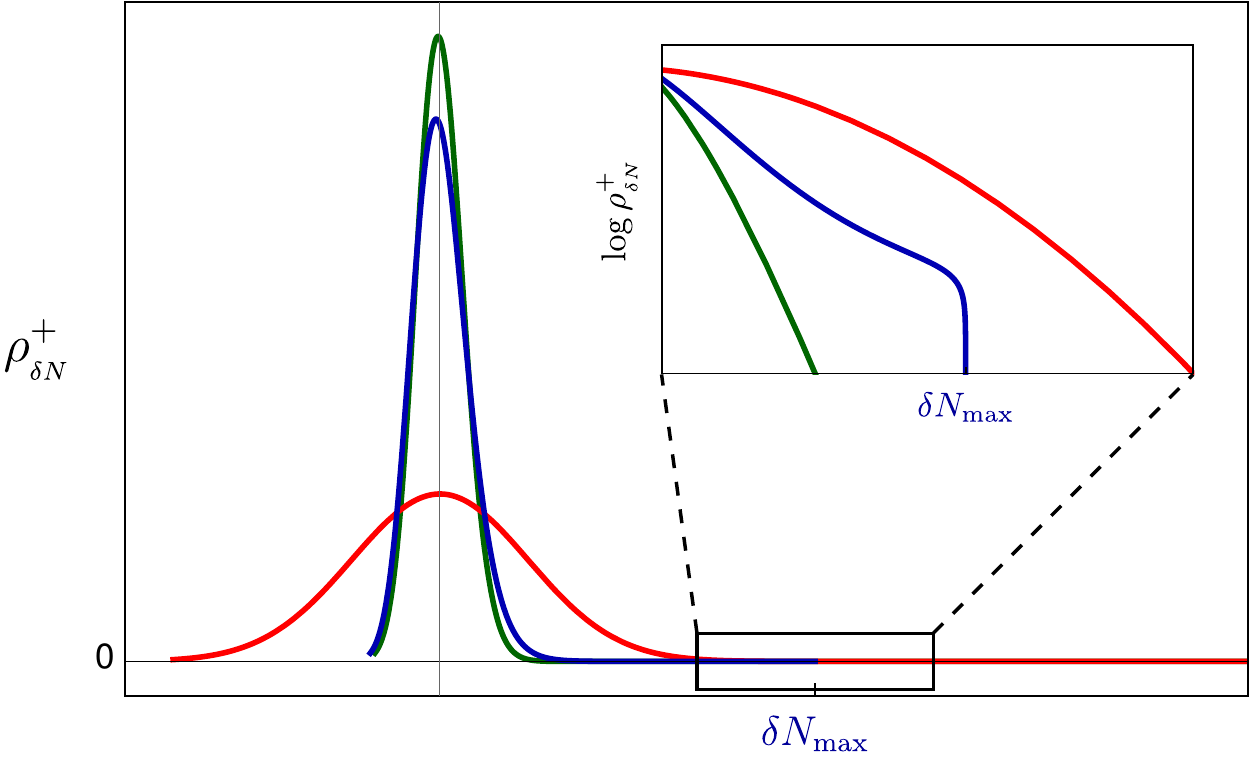}
		\end{center}
		\caption{  The schematics of the $\rho_{_{\delta \N}}^+$ piece of PDF for the three cases (i), (ii) and (iii) (explained in the text) with the linear potential Eq.~\eqref{linear-V}. The blue tick depicts where the blue curve (i.e., the PDF for the case (iii)) is truncated. The inset plot illustrates the behavior of the three PDFs around the truncation point in logarithmic scale.  The $\rho_{_{\delta \N}}^-$ piece must be added to the left from where the PDFs are cropped. }
		\label{fig:linear-V-PDf}
	\end{figure}
	
	Depending on  the choice of parameters, there are three qualitatively different possibilities for the solutions described by Eq.~\eqref{phi(N)-linear}: (i) $\alpha>0$, $\bar\pi<0$ where the field simply rolls down the potential, (ii) $\alpha>0$, $\bar\pi>0$ where the field goes up before turning around and rolling down the potential, and (iii) $\alpha<0$, $\bar\pi<0$ where the field climbs up the potential. 
	
	The behavior of $\phi(n)$ is depicted in the left panel of Figure~\ref{fig:linear-V-deltaN} for the three cases at hand.  The first crossing of $\phi_c$ can be read from this figure, which can then be translated to a plot of $\delta\phi$ vs.\ $\delta\N$, shown in the right panel of Figure~\ref{fig:linear-V-deltaN}.  The asymptotic linear behavior of $\delta\phi$ vs.\ $\delta\N$, which we alluded to above (after Eq.~\eqref{phi_c-linear}), can clearly be seen here.  It can also be seen from the right panel of Figure~\ref{fig:linear-V-deltaN} that for cases (i) and (ii) the relation between $\delta\phi$ and $\delta\N$ is one-to-one.  Therefore, finding $\rho^+_{\delta\N}$ from $\rho_{\delta\phi}$ is done by a simple change of variable from $\delta\phi$ to $\delta\N$.  The resulting PDFs $\rho^+_{\delta\N}$ are shown in Figure~\ref{fig:linear-V-PDf}, where the Gaussian tails extend to the right where $\delta\N\to\infty$.  Note that, as mentioned earlier, we have opted not to show the Gaussian $\rho^-_{\delta\N}$, but in principle it has to be glued to the left of the curve where it is cropped.
	
	Case (iii) (where the field needs to climb up to reach $\phi_c$) is distinct in that if $\delta\phi$ is large enough, the initial velocity $\bar\pi$ is insufficient to get the field up to $\phi_c$. The maximum allowed field perturbation $\delta\phi_{\rm max}$ (corresponding to the dot-dashed curve in the left panel of Figure~\ref{fig:linear-V-deltaN} corresponds to a maximum $\delta\N$ given by
	\begin{equation}
	\delta\N_{\rm max} = \frac13 \log \frac{\bar\pi+\alpha}{\alpha} - \bar \N_c\,  ,
	\end{equation}
	which may be derived by using Eq.~\eqref{dphi_N_lin} and solving $\delta \phi'( \delta\N_{\rm{max}})=0$  for $\delta \N_{\rm{max}}$.
	It can then be seen from the right panel of Figure~\ref{fig:linear-V-deltaN} (the blue curve) that the relation between $\delta\phi$ and $\delta\N$ ceases to be one-to-one beyond $\delta\N_{\rm max}$.  The points on the curve beyond the one-to-one regime correspond to the trajectories that start from $\bar \phi+\delta \phi$ with $\delta \phi < \delta \phi_{\rm max} $, climb up the potential to pass $\phi_c$, reach some maximum height and return to reach $\phi_c$ for the second time. Thus, the points beyond $\delta \N_{\max}$ (blue, dot-dashed curve in the right panel of Figure~\ref{fig:linear-V-deltaN}) correspond to the second crossing of the end point $\phi_c$ which must be excluded. This exclusion implies that we have a truncated PDF which vanishes for $\delta\N > \delta\N_{\rm max}$.  To investigate the behavior of the PDF in the left neighborhood of $\delta\N_{\rm max}$, we note that  $d\delta\phi/d\delta\N$ vanishes at $\delta\N_{\rm max}$; but this is precisely the Jacobian factor appearing in $\rho_{_{\delta \N}}^+$.  We therefore find that the PDF vanishes as $\delta\N$ approaches $\delta\N_{\rm max}$ from left.
	It is to be noted that this is a continuous truncation, i.e., 
	$\displaystyle \lim_{\delta\N\to\delta\N_{\rm max}\pm} \rho_{_{\delta \N}}$ is zero from \textit{both} sides (although the higher derivatives of $\rho_{_{\delta \N}}$ are not continuous).  We therefore call it a second-order truncation (in analogy with the second-order phase transitions), to contrast with the discontinuous truncation that we will encounter in Sec.~\ref{sec:barrier}.
	
	As a final remark, notice that while classically reaching $\phi_c$ is  forbidden for fluctuations larger than $\delta \phi_{\max}$, stochastic effects---in principle---may open up new possibilities. However, since the classical behavior rapidly moves the field away from $\phi_c$, it is very unlikely that the stochastic effects significantly alter our classical picture, except for a negligible fraction of possible fluctuations very close to (and slightly larger than) $\delta \phi_{\max}$.

	\subsubsection{Quadratic potential}
	\label{sec:quadratic}
	
	We now consider the special case $\alpha=0$ of the potential \eqref{general-V2}. To simplify our results, we assume that the field starts from the bottom ($\bar\phi=0$) of the quadratic potential $V=V_0(1+\frac12\beta\phi^2)$.
	
	The first distinct feature of the quadratic potential, compared to the linear one, is the shape of its tail.  As we saw in Sec.~\ref{sec:linear}, the linear potential leads to a Gaussian tail (if not truncated at some $ \delta \N_\text{max}$).  For the quadratic potential, we have to inspect the large $\N$ limit of Eq.~\eqref{delta-phi-zeta-general} (with $\phi_a=0$).  Assuming $\beta <3/4$, we find 
	\begin{equation}
	\delta\phi \approx \frac{2\Gamma \phi_c}{1+\Gamma} e^{\frac32 (1-\Gamma)\N}\, , \qquad \text{for large $\N$} , 
	\end{equation}
	which is indicative of a double-exponential tail,
	\begin{equation}
	\rho_{_{\delta \N}} \sim \exp \left[ -c \, e^{3(1-\Gamma) \delta \N} \right]\, , \qquad \text{for large $\delta \N$} , 
	\end{equation}
	where $ c$ is some $\delta \N$-independent constant.  
	
	\begin{figure}
		\centering
		\includegraphics[scale=.7]{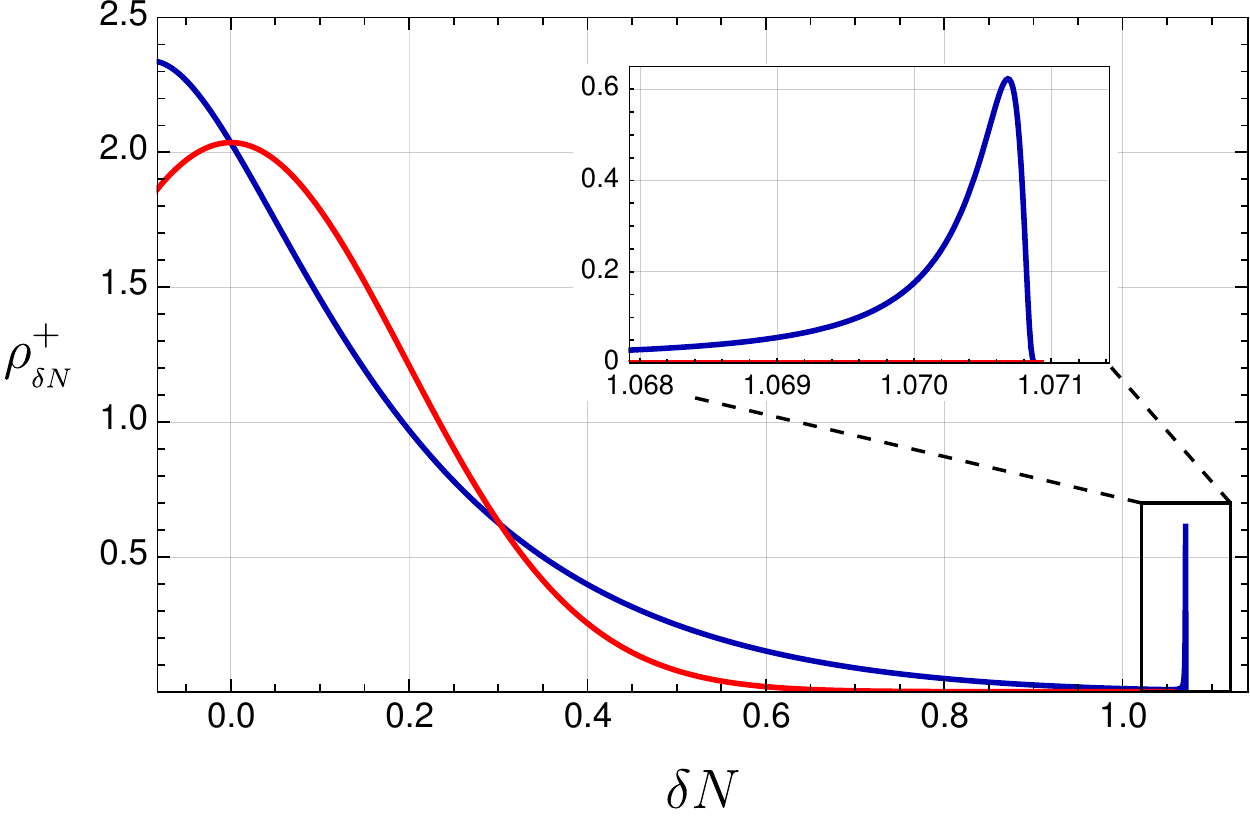}
		\caption{A PDF for a quadratic potential (Eq.~\eqref{general-V2} with $\alpha=0$) featuring two peaks.  The red curve is a Gaussian PDF obtained from the naive linear approximation. In the inset panel, we zoom in on the area around the second peak. The other parameters are $\bar\phi=0$, $\bar\pi = -5.9\times10^{-5}$, $\tilde\Gamma = 1.2894$, $V_0=9.6 \times 10^{-9}$ and $\phi_c=-5 \times 10^{-2}$ leading to $\sigma_{\delta\phi} \simeq 9\times10^{-6}$ and $\bar N_c \simeq 0.0823$.}
		\label{fig:2-peak}
	\end{figure}
	
	A second novel feature of the quadratic potential appears when we consider  $\beta>3/4$.  In this case, Eq.~\eqref{delta-phi-zeta-general} changes to
	\begin{equation}\label{delta-phi-zeta-quadratic}
	\delta\phi = \frac{\displaystyle \tilde \Gamma \phi_c e^{\frac{3\N}{2}}  - \frac23 \bar\pi \sin \frac{3\tilde \Gamma\N}{2}}{\displaystyle \tilde \Gamma \cos \frac{3\tilde \Gamma\N}{2} + \sin \frac{3\tilde \Gamma\N}{2} },
	\end{equation}
	where $\tilde \Gamma \equiv \sqrt{\frac{4\beta}{3}-1}$ and recall that we have set $\bar \phi = \phi_a =0$. Crucially, the denominator of  this equation vanishes if $\N$ satisfies $\tan (3\tilde \Gamma\N/2) = -\tilde \Gamma$.  This determines $\delta \N_{\max}$ so, again, we have truncation of the PDF. However, this has a very different physical reason as compared to the truncation in the linear potential, discussed in Sec.~\ref{sec:linear} (case (iii)). There, the truncation occurs since for sufficiently large fluctuations the field is unable to reach $\phi_c$. Here, the slope of the potential is so high that one cannot arbitrarily increase the $e$-folds to reach $\phi_c$, by increasing $\delta \phi$. This also implies that $\delta \phi_{\rm max} \to \infty$ and $p_r=1$.
	
	Another  novel feature of this model is the appearance of two peaks in the PDF, albeit with suitably chosen set of parameters, as shown in Figure~\ref{fig:2-peak}.   They are chosen such that the secondary peak is beyond $\zeta=1$ and contributes to the PBH abundance. The second peak is  a result of the non-linear relation between $\delta \phi$ and $\delta \N$ being so complex that different factors with different behaviors  contribute to the PDF of $\delta \N$, allowing for a  temporary enhancement.
	More explicitly, notice that around $\delta \N_{\max}$, Eq.~\eqref{delta-phi-zeta-quadratic} can be approximated by $\delta \phi \simeq \lambda \, (\delta \N-\delta \N_{\max})^{-1}$ where 
	\begin{equation}
	\label{gamma}
	\lambda \equiv \frac{4 \bar \pi }{9 \left(1+\tilde\Gamma ^2\right)} - \frac{2  \phi_c }{3 \sqrt{1+\tilde\Gamma ^2}}\, e^{3 \N_{\max}/2}\, ,
	\end{equation}
	where $\N_{\max} = \frac{2}{3\tilde \Gamma} (\pi-\tan^{-1}\tilde \Gamma)$.
	This approximation leads to
	\begin{equation}
	\label{pdf_quadratic_approx}
	\rho_{_{\delta \N}} \simeq  \frac{|\lambda|}{\sqrt{2\pi} \delta^{2} \,  \sigma_{\delta \phi} }\, e^{-\frac{\lambda^2 }{2 \delta^{2}\, \sigma_{\delta \phi}^2}} \qquad   \text{for \, $\delta \N \sim \delta \N_{\max}$} \, ,
	\end{equation}
	where $\delta \equiv \delta \N-\delta \N_{\max}$.
	This expression clearly has a peak at 
	\begin{eqnarray}
	\delta \N_{\text{peak}}=\delta \N_{\max}-\dfrac{|\lambda|}{\sqrt{2} \sigma_{\delta \phi}} \, , 
	\end{eqnarray}
	with the height
	\begin{equation}
	\rho_{_{\delta \N}}(\delta \N_{\text{peak}}) \simeq \sqrt{\frac2\pi}\, \dfrac{\sigma_{\delta \phi}}{e |\lambda|}. 
	\end{equation}
	By reducing the value of $|\lambda|$ (e.g., by fine tuning the parameters  so that the  two terms of Eq.~\eqref{gamma} nearly cancel) one can raise the peak to high values. The behavior of PDF according to Eq.~\eqref{pdf_quadratic_approx} also shows that  the truncation at $\delta \N_{\max}$ is second-order  (i.e., the PDF goes to zero from both sides).
	
	In Table~\ref{tab:quadratic} we present the values of $\mathcal{L}_3 /\mathcal{L}_2$, $\cal S$ and  $\beta_\zeta$  for this model which are, respectively, the strength of self-interaction of $\delta \phi$ (Eq.~\eqref{L3/L2}),  the significance of stochasticity (Eq.~\eqref{S})  and the probability that $\zeta>1$ can be realized (Eq.~\eqref{beta_def}). Furthermore, $\beta_{\zeta}^{(1)}$ is the prediction of the linear theory $\delta \N\simeq \N'(\bar \phi) \delta \phi$ (leading to a Guassian PDF). The difference between $\beta_{\zeta}$ and $\beta_{\zeta}^{(1)}$ is both because of the non-trivial behavior of the full PDF in the intermediate regime and its second peak near the truncation point.
	
	\begin{table}[]
		\center
		\caption{Different quantities for the quadratic potential with the same parameters as in Figure~\ref{fig:2-peak}. }
		\begin{tabular}{|c|c|c|c|}
			\hline
			$\beta_\zeta^{(1)}$              & $\beta_{\zeta}$ & $\mathcal{L}_3 /\mathcal{L}_2$ & $\mathcal{S}$         \\ \hline
			$6.0 \times 10^{-8}$ &   $0.00041$     & $7.1 \times 10^{-9}$          & $9.7 \times 10^{-6}$ \\ \hline
		\end{tabular}
		\label{tab:quadratic}
	\end{table}
	
	\subsection{Power-law tail}
	\label{sec:power-law} % $p$-potential; extension of the previous work
	
	In this section we propose a class of potentials that lead to a power-law tail for $\rho_{_{\delta \N}}$, i.e., one that falls like a negative power of $\delta \N$.  We will present a suggestive argument and eventually our  numerical calculation will prove that the tail is indeed power-law. We also provide further evidence for this in Appendix~\ref{app:V-p}. This section is  a generalization of Ref.~\cite{Hooshangi:2021ubn} where we had studied a simple model yielding a PDF  with a tail behaving like $1/\delta \N^2$.
	
	In order to specify a model that shows the desired behavior of the tail, we start by requiring that the Gaussian fall-off of PDF due to $\rho_{\delta\phi}$  in Eq.~\eqref{PDF-zeta-vs-delta-phi} terminates for sufficiently large values of $\delta \N$. This can be achieved if $\delta\phi(\N)$ approaches a finite constant (which we call $\delta\phi_\text{max}$) in the  large-$\N$ limit. One possibility to fulfill our requirement is to demand  the following asymptotic behavior for $\delta\N$
	\begin{equation}\label{deltaN-deltaphi}
	\delta\N \sim \left\{ \frac{1}{\gamma (\delta\phi_\text{max} - \delta\phi)} \right\}^{1/p}\, ,
	\end{equation}
	where $p$ and $\gamma$ are positive numbers.  Setting aside the technicalities of Sec.~\ref{sec:deltaN_large}, the resulting PDF is, according to Eq.~\eqref{PDF-zeta-vs-delta-phi}, proportional to  $|d\delta \phi/d\delta\N | \, \rho_{\delta\phi}(\delta\phi)$. The second factor  is  required to approach a constant for large $\delta \N$ while the first factor, according to Eq.~\eqref{deltaN-deltaphi}, yields a power-law decay of the PDF behaving like $1/\delta \N^{1+p}$.  This  is indeed practically heavy-tailed---as defined below Eq.~\eqref{D}---with ${\cal D} \to  (1+p)/\delta \N \to 0$ for large $\delta \N$ (where, $\cal D$ is defined in Eq.~\eqref{D}).  Note that $\delta\phi_{\rm max}$ here is also  consistent with Sec.~\ref{sec:deltaN_large} where we employed $\delta\phi_{\rm max}$ to denote the maximum possible fluctuation.  
	
	Next we need to show that there exists a potential that supports this solution.  One way to find the potential is to know a trajectory $\phi(n;\phi_0)$, or equivalently its inverse $n(\phi;\phi_0)$,\footnote{In our notation, $\phi(n;\phi_0)$ is the value of the field $n$ $e$-folds after it was at $\phi_0$.  Also, $n(\phi;\phi_0)$ is the number of $e$-folds required to reach $\phi$ from the initial value $\phi_0$.  We have suppressed the dependence on the initial velocity $\pi_0$ since we work with a fixed $\pi_0 = \bar\pi$ and ignore $\delta\pi$ perturbations.}
	However, a mere knowledge of $\N(\phi) = n(\phi_c;\phi)$ (let alone its asymptotic behavior) is insufficient to uniquely determine $n(\phi;\phi_0)$.  Therefore, strictly speaking, we have to make a guess. However, inspired by Eq.~\eqref{deltaN-deltaphi}, we naturally adopt 
	\begin{equation}\label{N_phi}
	n(\phi; \bar\phi) \sim \left\{ \frac1{\gamma \left( \phi - \bar\phi \right)} \right\}^{1/p}.
	\end{equation}
	Assuming that $\phi$ evolves monotonically, one can convert the Klein-Gordon  equation ~\eqref{KG} to the following equation for $n(\phi;\phi_0)$:
	\begin{equation}\label{KG-inf}
	\frac{V_{,\phi}}{V} = -\frac{1}{n_{,\phi}} + \frac{n_{,\phi\phi}/n_{,\phi}}{3n_{,\phi}^2 - \frac12}.
	\end{equation}
	Plugging Eq.~\eqref{N_phi} into Eq.~\eqref{KG-inf}, a simple integration results in
	\begin{equation}
	V(\phi) \, \sim \, V_0 \left[ 1 - \frac{p^2}{6} \gamma^{2/p} \left( \phi - \bar\phi \right)^{2+2/p} \right] \exp \left[ \frac{p^2}{2p+1} \gamma^{1/p} \left( \phi - \bar\phi \right)^{2+1/p} \right].
	\end{equation}
	We will work in the regime that $\gamma|\phi-\bar\phi|^{p+1} \ll 1$ and $\gamma|\phi-\bar\phi|^{2p+1} \ll 1$ (the first of these two conditions corresponds to the de~Sitter limit $\epsilon = \frac12 N_{,\phi}^{-2} \ll 1$).  Therefore, we take the final form of our potential to be 
	\begin{equation}\label{V-p}
	V(\phi) = V_0 \left[ 1 + \frac{p^2}{2p+1} \gamma^{1/p} \left( \phi - \bar\phi \right)^{2+1/p} \right].
	\end{equation}

	\begin{figure}
		\centering
		\includegraphics[width=.45\textwidth]{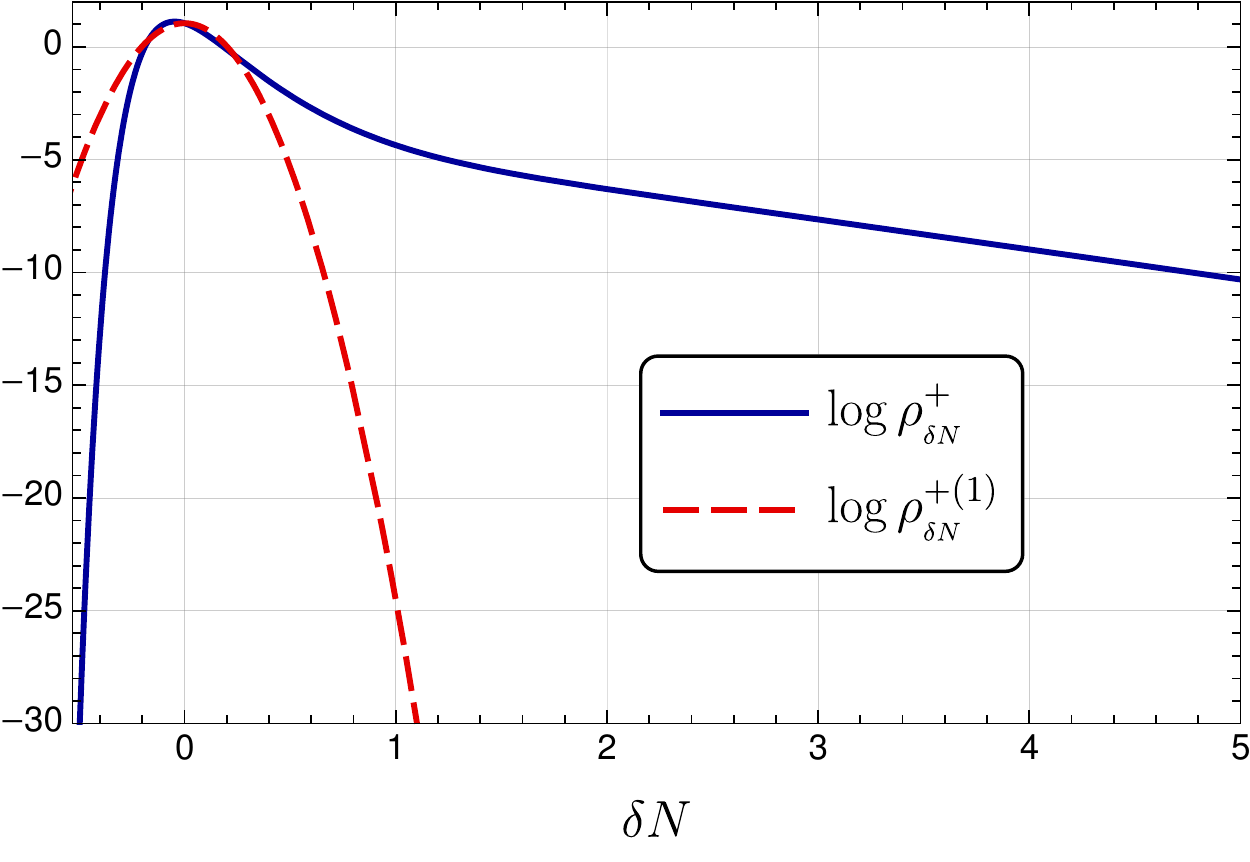}
		\hspace{.5cm}
		\includegraphics[width=.45\textwidth]{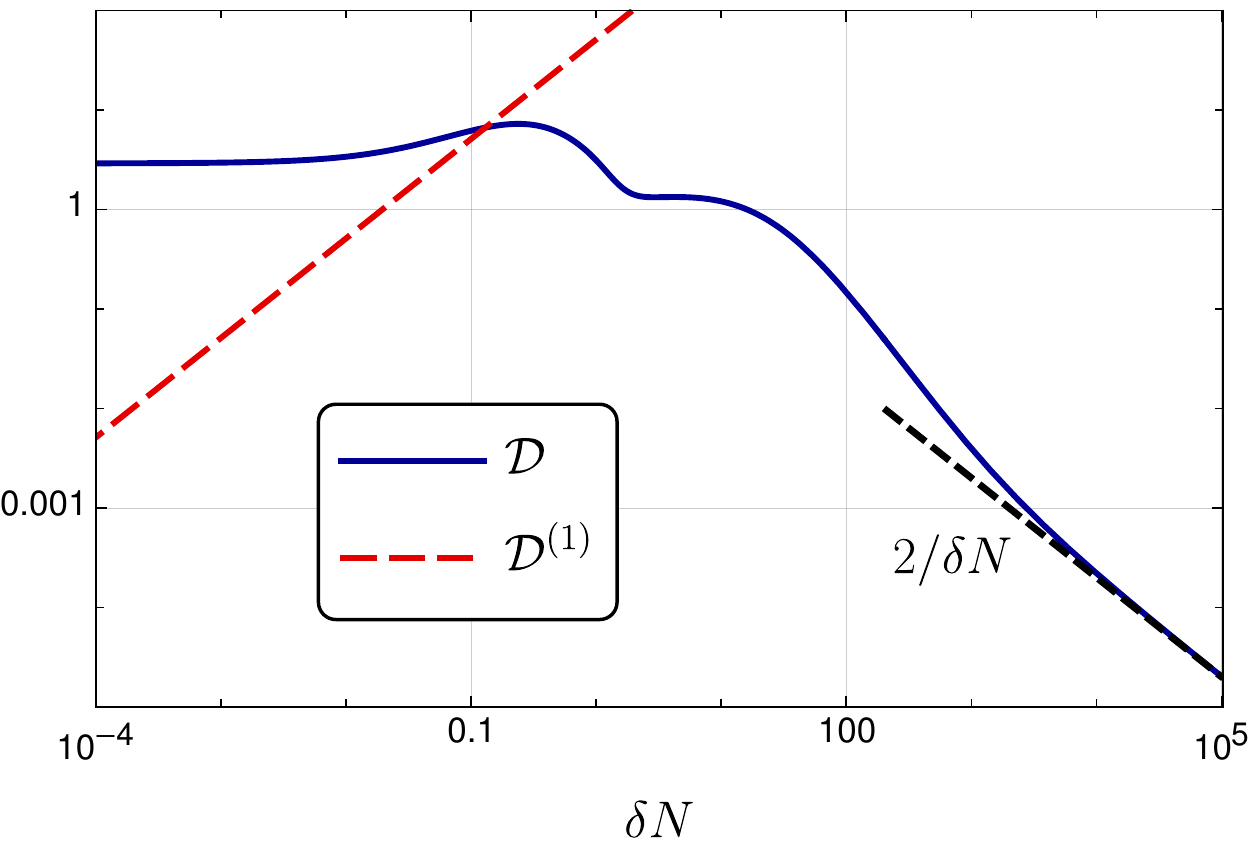}
		\caption{$\rho_{_{\delta \N}}^+$ (left) and the heaviness measure $\cal D$ (right), for the potential~\eqref{V-p} with $p=1$, $\bar\phi=0$, $V_0=10^{-9}$, $\gamma=2 \times 10^3$, $\bar\pi=-10^{-4}$, $\phi_c=-2.67\times 10^{-5}$. These parameters yield $\bar N_c \simeq 0.53$. The blue curve is the full nonlinear result, while the dashed red curve is the result of the linear approximation.}
		\label{fig:p=1}
	\end{figure}
	
	\begin{figure}
		\centering
		\includegraphics[width=.45\textwidth]{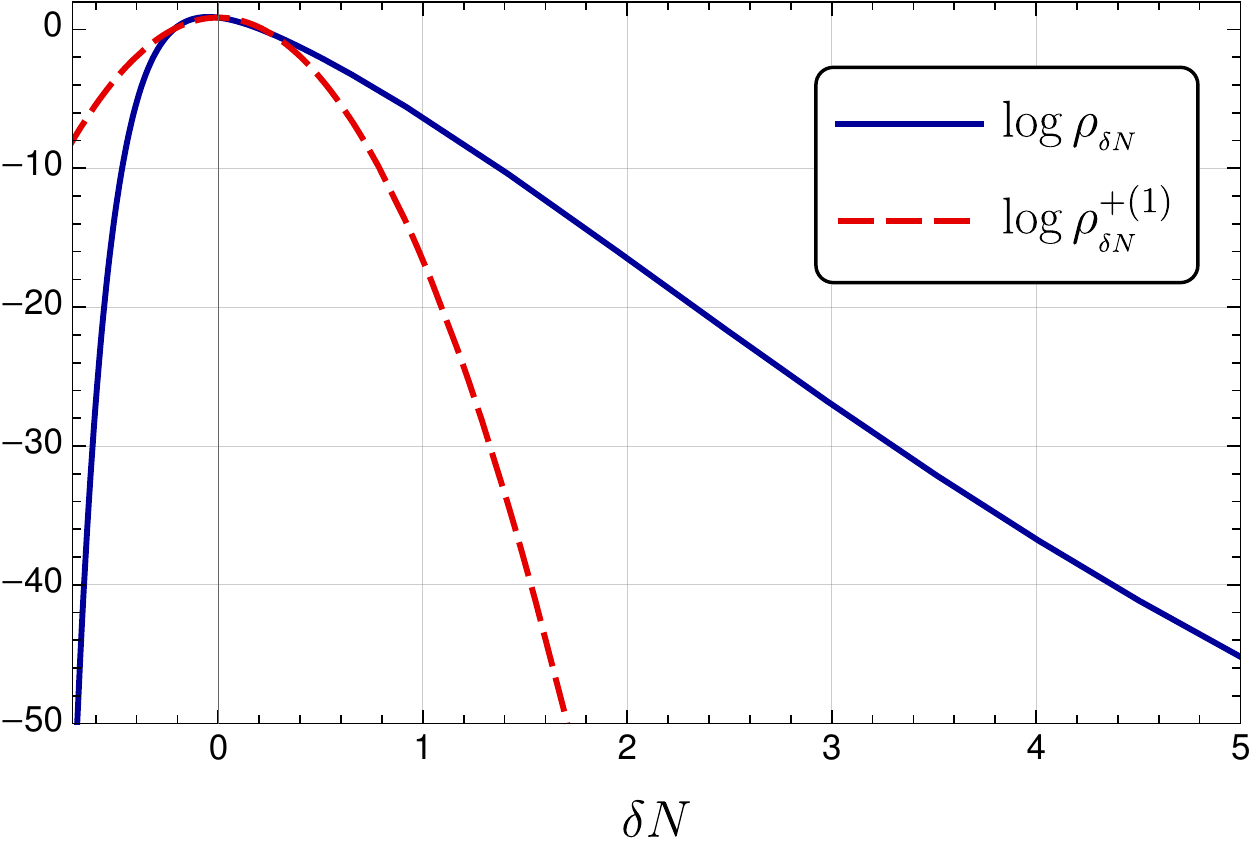}
		\hspace{.5cm}
		\includegraphics[width=.45\textwidth]{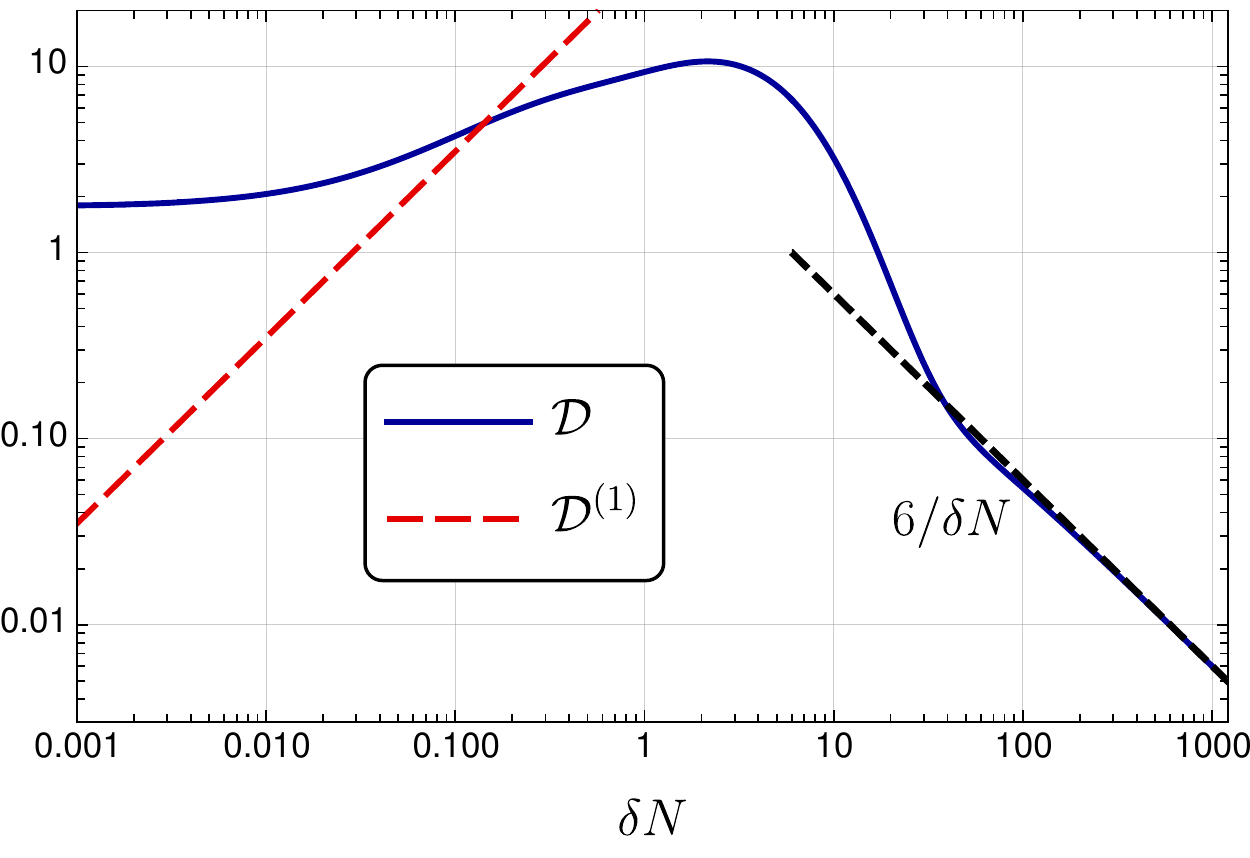}
		\caption{Same as the previous figure, but for $p=5$, $\bar\phi=0$, $V_0=10^{-9}$, $\gamma=10^{-1}$, $\bar\pi=-10^{-4}$ and $\phi_c=3.2 \times 10^{-5}$, resulting in $\bar N_c\simeq 0.71$ .}
		\label{fig:p=5}
	\end{figure}
	
	\begin{figure}
		\centering
		\includegraphics[width=.45\textwidth]{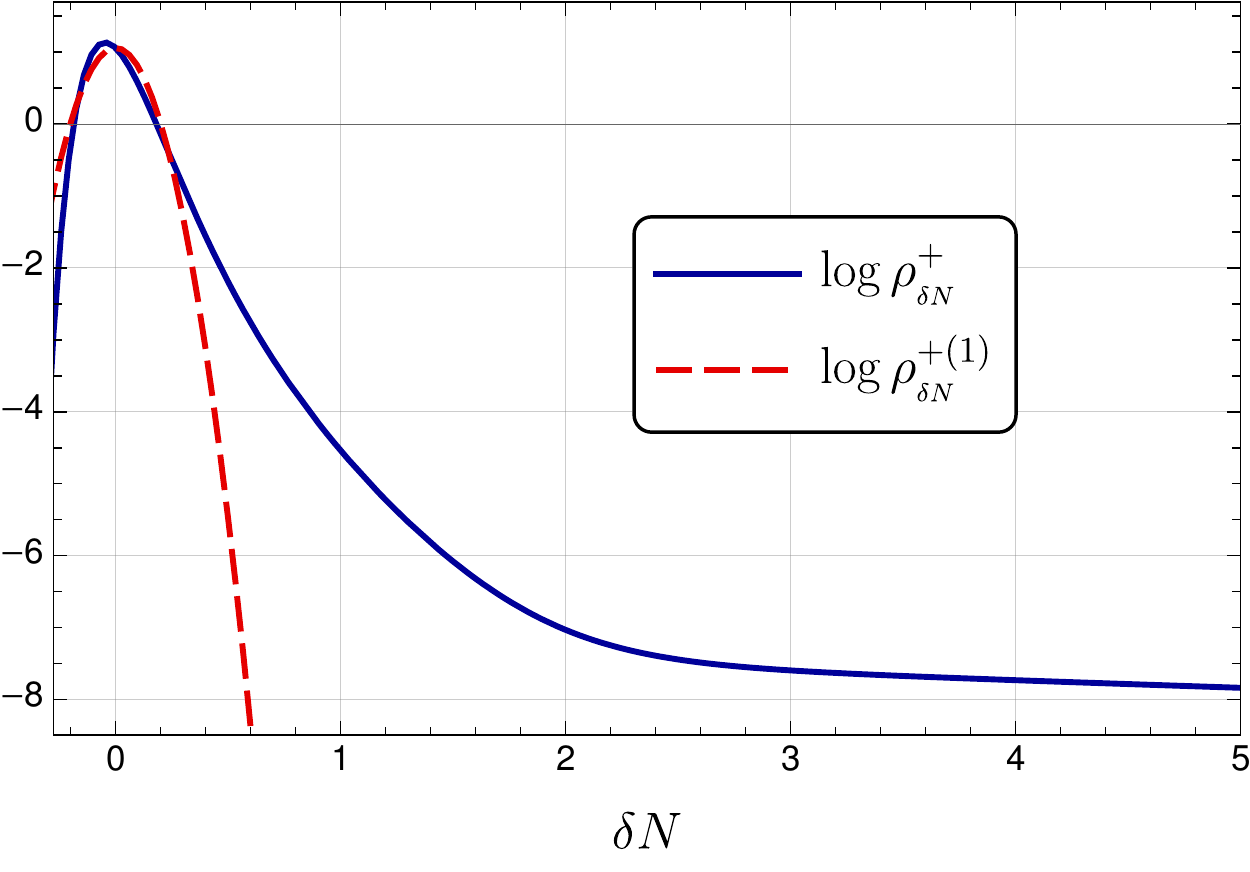}
		\hspace{.5cm}
		\includegraphics[width=.45\textwidth]{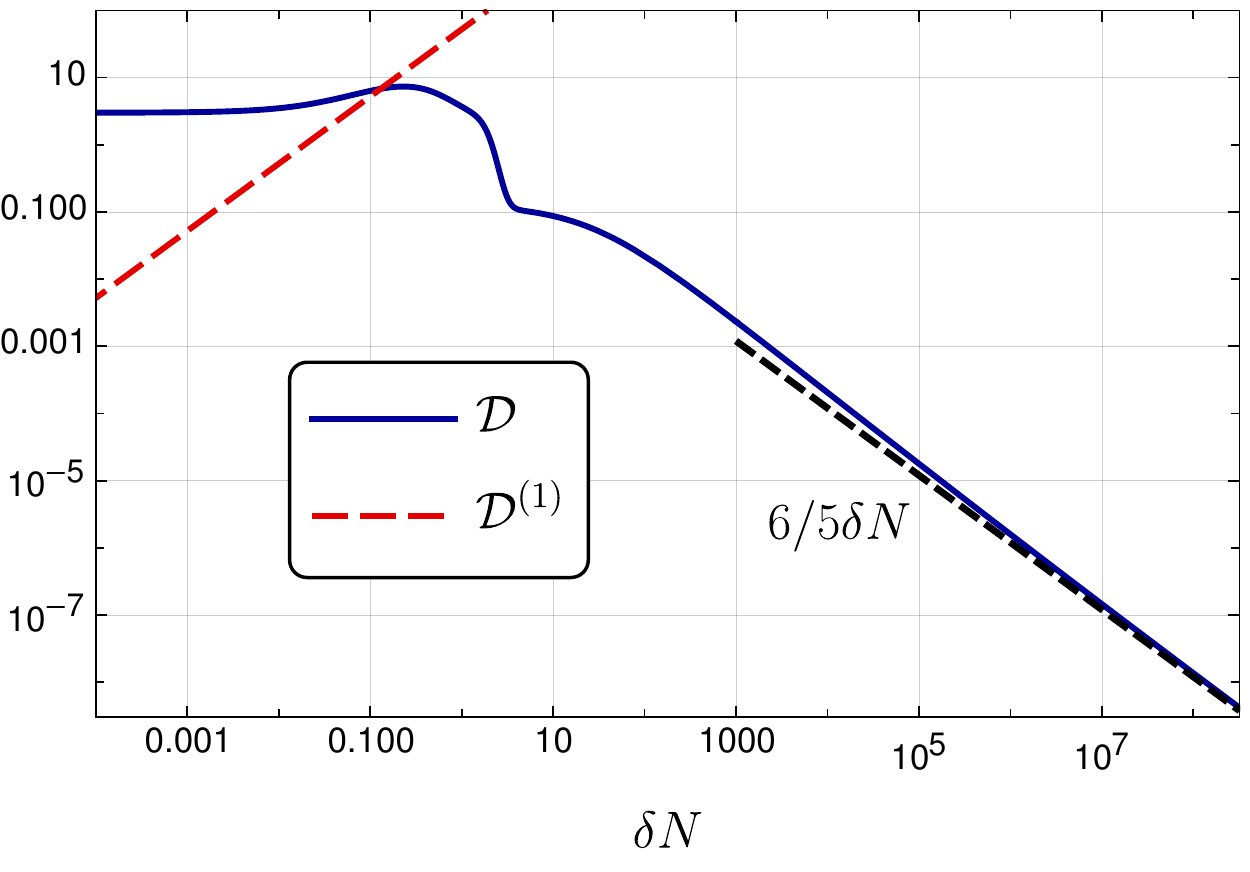}
		\caption{Same as the previous figure, but for $p=1/5$, $\bar\phi=0$, $V_0=10^{-9}$, $\gamma=6 \times 10^4$, $\bar\pi=-4.8 \times 10^{-5}$ and $\phi_c=-9 \times 10^{-6}$ which yield $\bar N_c \simeq 0.28$.}
		\label{fig:p=1/5}
	\end{figure}
	
	We have verified numerically that this $V(\phi)$ indeed leads to a power-law tail decaying like $1/\delta \N^{1+p}$, albeit for sufficiently large $\delta \N$.\footnote{One can choose a different set of parameters so that the PDF converges to the power-law prediction much  earlier. Typically, the price to pay is an increase in the significance of stochasticity, which we tried to avoid in the examples we present in this paper.}  This is shown in Figures~\ref{fig:p=1}, \ref{fig:p=5}, and \ref{fig:p=1/5}, for $p=1$ (also studied in Ref.~\cite{Hooshangi:2022lao}), $p=5$, and $p=1/5$, respectively, where the asymptotic behavior of the heaviness measure $\cal D$ perfectly matches our analytic expectation for the tail. In Table.~\ref{tab:p} we confirm the validity of our basic assumptions, discussed in Sec.~\ref{sec:assumptions}, and also compare $\beta_{\zeta}$ from the PDFs for the three  values of $p$ with the Gaussian counterpart.

	In this model, we have chosen the parameters such that $|\delta \phi_c|$ is sufficiently larger than $\sigma_{\delta \phi}$  (which is equivalent to sufficiently large $\bar N_c$)  so that the role of $\rho_{_{\delta \N}}^-$ becomes insignificant both for the normalization and for the computation of $\beta_\zeta$. On the other hand, since the decay of the PDF at its tail is so slow---especially for small values of $p$---the constant shift that relates $\delta \N$ to $\zeta$, according to Eq.~\eqref{zeta_def}, does become important. In particular, the case with $p=1/5$ has $\langle \delta \N \rangle\simeq  2.26$ which leads to a significant shift.
	
	\begin{table}[]
		\center 
		\label{tab:p}
		\caption{This table illustrates the validity of our assumptions and shows the values of $\beta_{\zeta}$ for the potential Eq.~\eqref{V-p} and for the three values of $p$. For comparison, we also present  $\beta_\zeta^{(1)}$ from the linear theory. For $p=5$, instead of $\mathcal{L}_3/\mathcal{L}_2$ according to Eq.~\eqref{L3/L2}, we used Eq.~\eqref{L3/L2_nonanalytic}.}
		\begin{tabular}{c|c|c|c|c|}
			\cline{2-5}
			& $\beta_\zeta^{(1)}$ & $\beta_\zeta$      & $\mathcal{L}_3/\mathcal{L}_2$ & $\mathcal{S}$ \\ \cline{2-5} \noalign{\vspace{0.4ex}} \hline
			\multicolumn{1}{|c||}{$p=1$}    & $4.4 \times 10^{-13}$    & $0.0058$ & $0.029$                     & $0.11$    \\ \hline
			\multicolumn{1}{|c||}{$p=5$}   & $1.7 \times 10^{-9}$     & $0.00015$  & $0.14$                      & $0.090$      \\ \hline
			\multicolumn{1}{|c||}{$p=1/5$} & $2.4 \times 10^{-13}$    & $0.0068$  & $0.22$                      & $0.34$      \\ \hline
		\end{tabular}
	\end{table}
%	
%		\begin{table}[]
%		\center 
%		\label{tab:p}
%		\caption{This table illustrates the validity of our assumptions and shows the values of $\beta_{\zeta}$ for the potential Eq.~\eqref{V-p} and for the three values of $p$. For comparison, we also present  $\beta_\zeta^{(1)}$ from the linear theory. \ma{For $p=5$, instead of $\mathcal{L}_3/\mathcal{L}_2$ according to Eq.~\eqref{L3/L2}, we used Eq.~\eqref{L3/L2_nonanalytic}.}}
%		\begin{tabular}{|c||c|c|c|c|}
%			\hline
%			& $\beta_\zeta^{(1)}$ & $\beta_\zeta$      & $\mathcal{L}_3/\mathcal{L}_2$ & $\mathcal{S}$ \\ \hline \hline
%			$p=1$   & $4.4 \times 10^{-13}$    & $0.0058$ & $0.029$                     & $0.11$    \\ \hline
%			$p=5$   & $1.7 \times 10^{-9}$     & $0.00015$  & $0.14$                      & $0.090$      \\ \hline
%			$p=1/5$ & $2.4 \times 10^{-13}$    & $0.0068$  & $0.22$                      & $0.34$      \\ \hline
%		\end{tabular}
%	\end{table}
	
	\subsection{Potential barrier}
	\label{sec:barrier} 
	
	It has been recently suggested by the authors of Ref.~\cite{Cai:2021zsp} that a tiny step in the inflationary potential can lead to dramatically different results in the abundance of the PBHs, from a large enhancement to a complete prohibition.  The basic idea put forward in Ref.~\cite{Cai:2021zsp} is that, in the presence of an upward step in the inflationary potential, there is a threshold velocity to climb the step, below which the inflaton gets stuck downstairs and cannot reach past the step.\footnote{In the special case of a sharp step with height $\Delta V$ in the potential located at $\phi_s$, the condition to climb the step is the availability of sufficient kinetic energy, i.e., $\frac12 H_s^2 \pi_s^2 > \Delta V$, where $\pi_s$ is the velocity just before the step.  If the field excursion between $\phi_s$ and the top of the step is nonzero, as in a smooth barrier, then it also has to overcome the Hubble friction along the way.  In either case, the existence of a threshold velocity is a direct consequence of the barrier.}  It is then argued that this threshold velocity corresponds to a threshold (upper bound) on the curvature perturbations beyond which the PDF vanishes, hence a truncation.
	
	A more careful analysis shows that the mere existence of a velocity threshold is not enough to obtain a PDF truncation.  Here we present a consistent implementation of the idea of ``truncation due to a barrier'' and elucidate some of the key assumptions and conditions for this to happen.
	
	\begin{figure}
		\begin{center}
			\includegraphics[scale=.7]{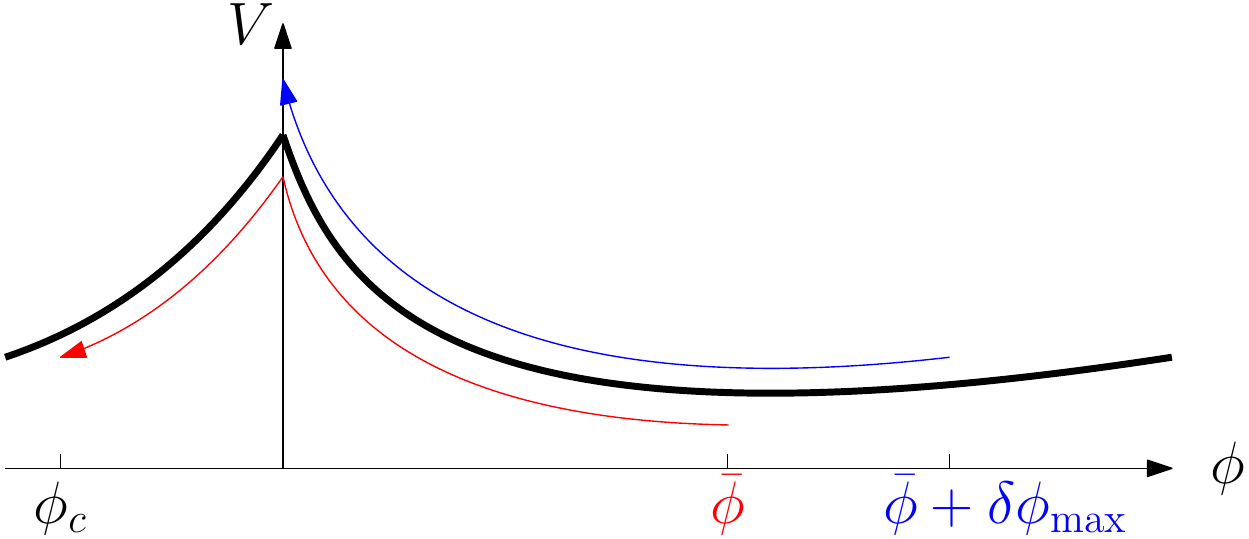}
		\end{center}
		\caption{A sketch of a potential that features truncation in the PDF.  The red and blue curves depict two trajectories with $\delta\phi=0$ and $\delta\phi = \delta\phi_{\rm max}$, respectively. }
		\label{fig:V-shutoff}
	\end{figure}
	
	To illustrate the setup, consider a field rolling down a potential $V(\phi)$ from right to left as in Figure~\ref{fig:V-shutoff}.  There is an upward barrier that the field has to climb up in order to reach the end point $\phi_c$.  In general, the barrier need not be a sharp step, so instead of constraining the velocity just before the step, we consider the initial field value $\bar\phi+\delta\phi$ and the initial field velocity $\bar\pi+\delta\pi$.  Keeping the initial velocity fixed (i.e., $\delta\pi=0$), the only parameter that controls whether or not the field has sufficient kinetic energy to pass the barrier is $\delta\phi$.  In other words, there exists a threshold $\delta\phi_{\rm max}$ on the initial field perturbations, beyond which the trajectories cannot reach $\phi_c$.  Note that we have to choose the initial values $\bar\phi$ and $\bar\pi$ so as to ensure the unperturbed trajectory does pass the barrier and reach $\phi_c$; otherwise we encounter eternal inflation and the fluctuations under consideration would be unobservable. The choice of $\bar \phi$ and $\bar \pi$ needs to be such that the climbing occurs in a non-attractor phase to allow for the super-horizon $\zeta$ to evolve and get affected by the step. On the other hand, the attractor velocity just below the step can be designed to be below the threshold velocity. This implies that for sufficiently large and positive fluctuations (corresponding to trajectories that start sufficiently far away from the step) the field cannot reach $\phi_c$ since its velocity has converged to the attractor value (which is below the threshold velocity). When $\delta\phi$ is negative (but smaller in magnitude than $\bar\phi-\phi_c$), $\bar\pi$ provides more than enough kinetic energy for traversing a shorter distance, so there is no negative threshold on $\delta\phi$. Hence, by continuity, there must be a point $\delta\phi_{\rm max}>0$ in between, starting from which the field will stop right at the top of the barrier and fall short of passing it. 
	
	We emphasize that even incorporating the above requirements does not necessarily amount to PDF truncation.  The existence of a threshold velocity only implies the existence of a $\delta\phi_{\rm max}$.  We have already seen in the working example of Sec.~\ref{sec:example_subtle} that $\delta\phi_{\rm max}$ can exist without leading to truncation.  To achieve truncation, it is essential that $\delta\phi_{\rm max}$ correspond to a \textit{finite} $\delta\N_{\rm max}$, such that $\rho_{_{\delta\N}}(\delta\N)=0$ for $\delta\N>\delta\N_{\rm max}$.  More formally, the limit
	\begin{equation}
	\lim_{\delta\phi\to\delta\phi_{\rm max}-} \delta {\N} = \delta \N_{\rm max}
	\end{equation}
	must be finite.\footnote{The right limit, namely $\lim_{\delta\phi\to\delta\phi_{\rm max}+} \delta {\N}$ does not exist, since fluctuations larger than $\delta\phi_{\rm max}$ must be excluded.  In other words, the trajectories that start from $\delta\phi > \delta\phi_{\rm max}$ never reach $\phi_c$, which is a direct consequence of having a threshold velocity.}  This implies that above the threshold velocity (equivalently, below the threshold perturbation $\delta\phi_{\rm max}$), not only does the field pass the barrier and reach $\phi_c$, but it does so in a finite time.  This is only possible when the potential is not smooth at the top of the barrier.  Indeed, if the barrier has a smooth peak at $\phi_s$ then, around the peak, we can approximate the potential by $V(\phi) \approx V(\phi_s) + \frac12 V''(\phi_s) (\phi-\phi_s)^2$. Neglecting the Hubble friction, the time required to reach the peak from a nearby point $\phi_i$ with the initial kinetic energy which is just enough to reach the peak is given by 
	\begin{equation}
	t = \int_{\phi_i}^{\phi_s} \frac{d\phi}{\sqrt{2(V(\phi_s)-V(\phi))}} = \int_{\phi_i}^{\phi_s} \frac{d\phi}{\sqrt{|V''(\phi_s)|} |\phi-\phi_s|},
	\end{equation}
	where  for the first equality we used the energy conservation (in the absence of friction) for the specific trajectory we considered. From the last equality it is evident that the time (and hence the $e$-folds) is unbounded which implies that there is no truncation in PDF (since arbitrarily large $\delta \N$ can be realized).  Therefore, a necessary condition to obtain a truncation  is sharpness of the peak of the barrier (to allow for a non-vanishing $V'(\phi_s)$, invalidating the above argument). The presence of the Hubble friction does not alter the above argument except a slight change in the critical trajectory that needs to be considered.
	
	Let us summarize what conditions are needed for the occurrence of PDF truncation as a result of a step in the potential (which do not seem to be spelled out in Ref.~\cite{Cai:2022erk}): (i) The unperturbed trajectory needs to be able to reach $\phi_c$ (otherwise, eternal inflation occurs which prohibits the fluctuations under consideration to be observable). (ii) The unperturbed trajectory must have a non-attractor velocity when it starts climbing up the step (otherwise, the super-horizon $\zeta$ freezes and there is no chance for the step to alter the statistics of $\zeta$). (iii) The attractor velocity just before the step (to which the trajectories with sufficiently large and positive $\delta \phi$ converge) needs to be insufficient to reach $\phi_c $ (otherwise, basically all trajectories do reach $\phi_c$, hence no truncation). (iv) The peak needs to be sharp so that $V'(\phi_s)$ does not vanish (otherwise, one can always find trajectories along which it takes arbitrarily long time to reach $\phi_c$---corresponding to arbitrarily large $\delta \N$---hence no truncation).

\begin{figure}
		\centering
		\includegraphics[width=.45\textwidth]{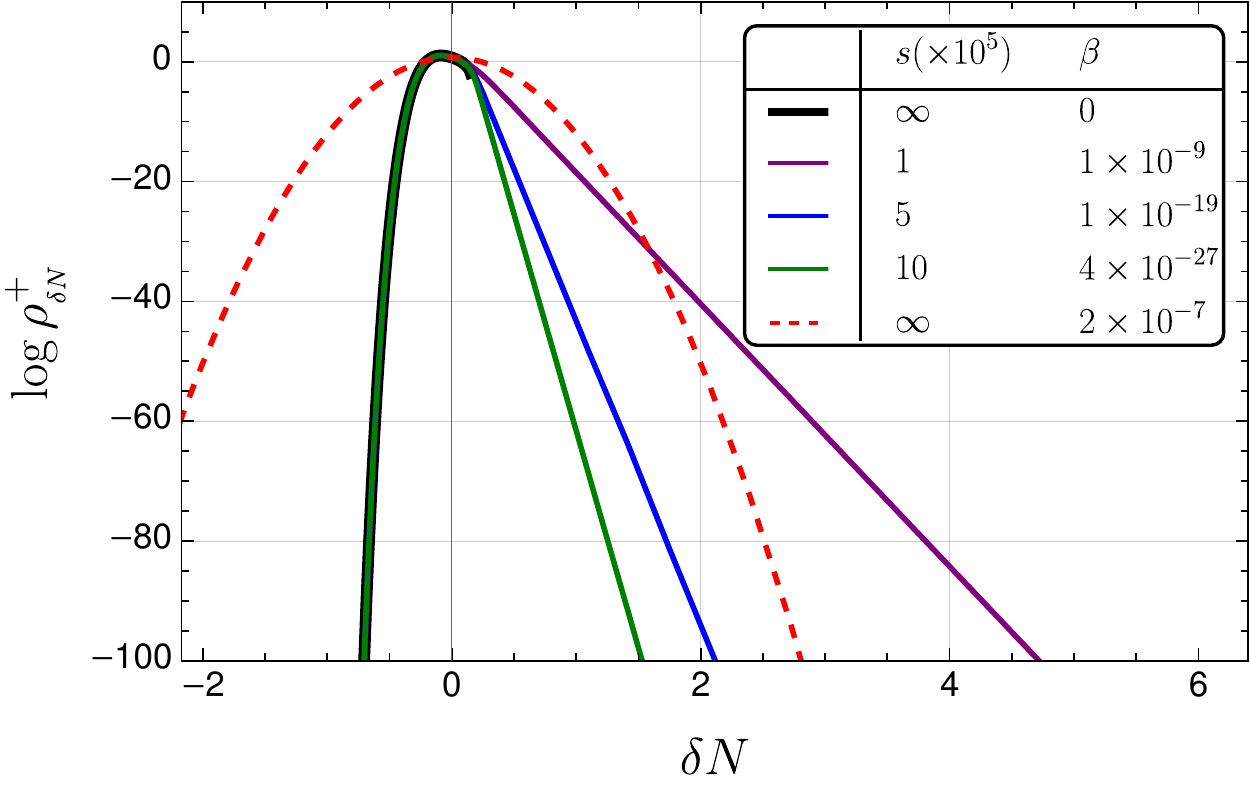}
		\hspace{.5cm}
		\includegraphics[width=.45\textwidth]{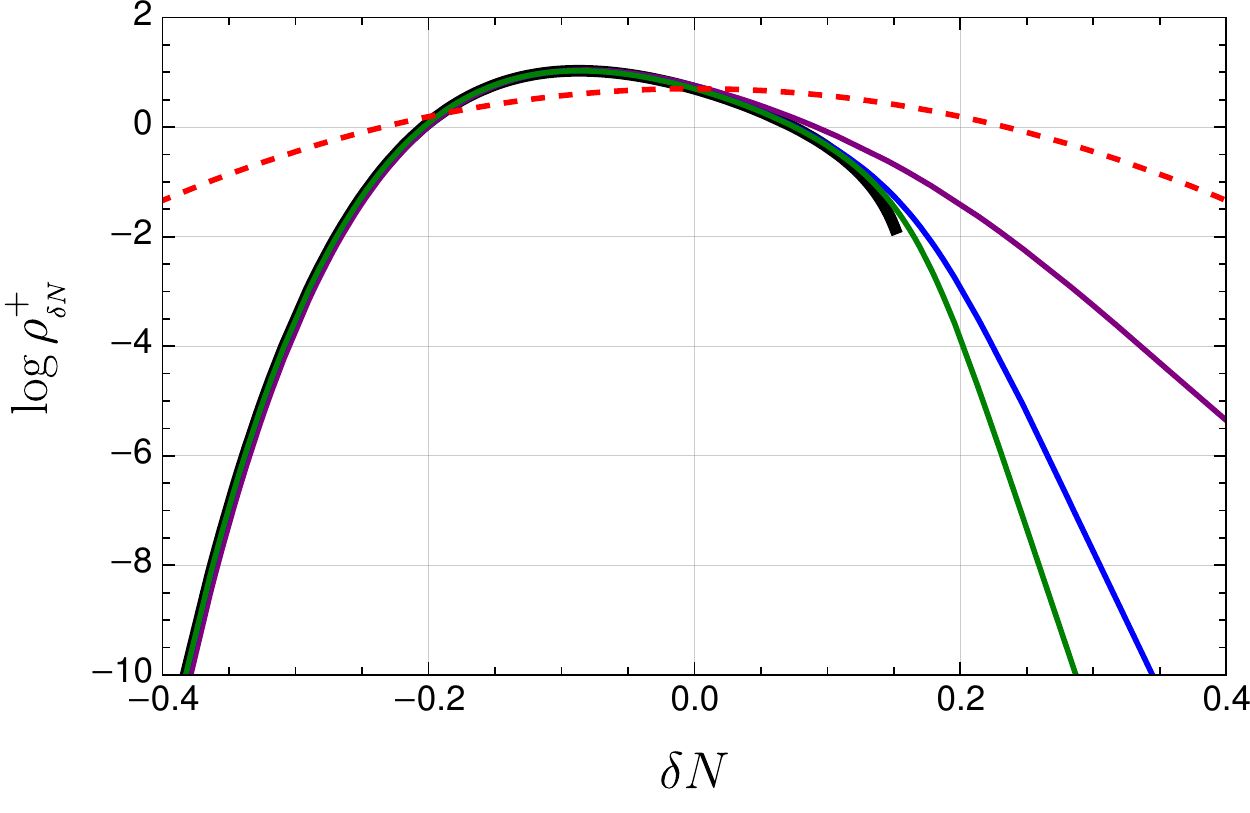}
		\caption{The PDF corresponding to the three-segment potential~\eqref{3-segment-V} (thick, black), its Gaussian approximation (red, dashed), and those of a few smoothed potentials with sharpness parameter $s$ (see Eq.~\eqref{sharpness}).  The corresponding $\beta_{\zeta}$s are also tabulated.  The right panel is a zoomed version of the left panel. The numerical values of the parameters are $V_1=1.18 \times 10^{-6}$, $\alpha_1=10^{-3}$, $\alpha_2=-10^{-3}$, $\alpha_3=10^{-2}$, $\phi_1=0$, $\phi_2=-1.532\times10^{-2}$, $\bar\phi=5.07424\times10^{-4}$, $\phi_c=-2.232\times10^{-2}$, $\bar\pi=-5.14923\times10^{-2}$. These lead to $\sigma_{\delta\phi}\simeq 10^{-4}$ and $\bar N_c\simeq 2.17$.  }
		\label{fig:shutoff-PDF-4}
	\end{figure}

\begin{figure}
		\centering
		\includegraphics[width=.45\textwidth]{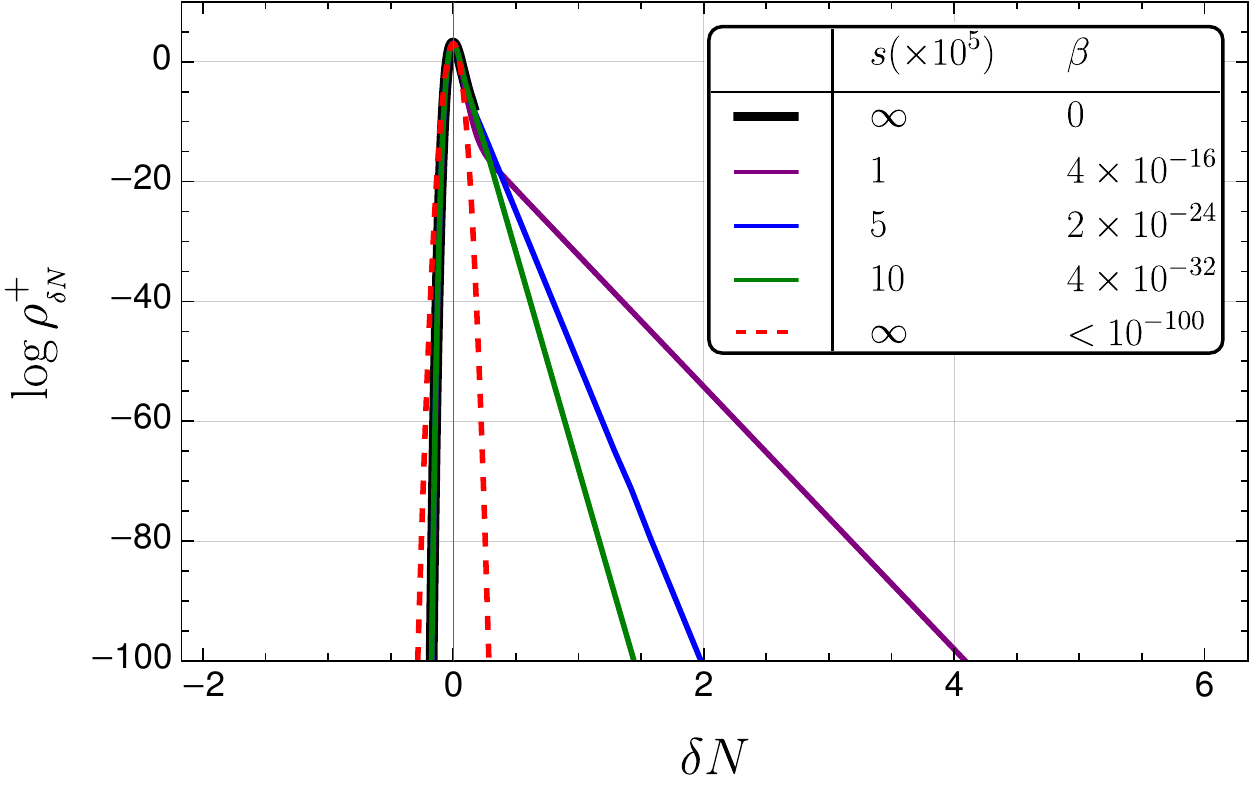}
		\hspace{.5cm}
		\includegraphics[width=.45\textwidth]{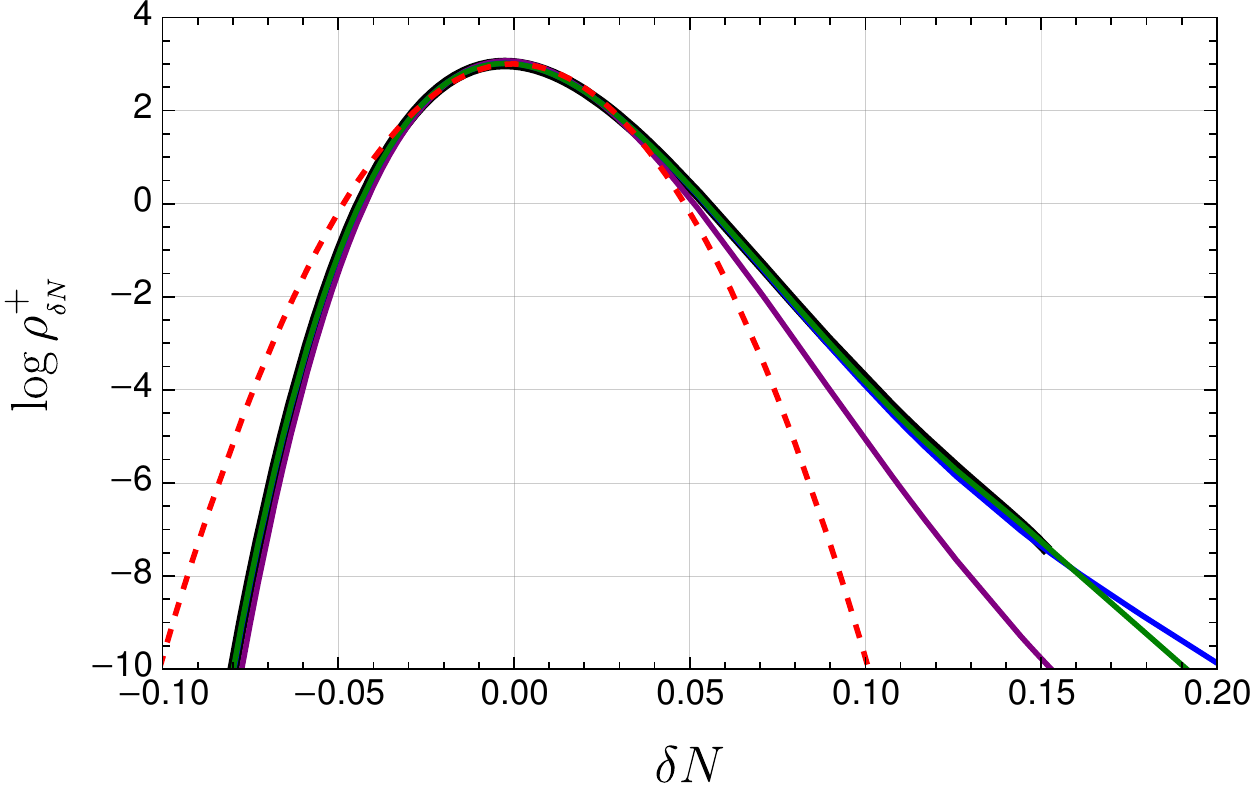}
		\caption{Same as the previous figure, but for $\sigma_{\delta\phi} \simeq 10^{-5}$ ($V_1=1.18 \times 10^{-8}$).}
		\label{fig:shutoff-PDF-5}
	\end{figure}

	As a concrete example, we consider a potential that is comprised of three linear pieces with three different slopes:
	\begin{equation}\label{3-segment-V}
	V = 
	\begin{cases}
	V_1 \left[ 1 + \alpha_1 (\phi - \phi_1) \right] & \phi > \phi_1, \\
	V_1 \left[ 1 + \alpha_2 (\phi - \phi_1) \right] & \phi_1 > \phi > \phi_2, \\
	V_2 \left[ 1 + \alpha_3 (\phi - \phi_2) \right] & \phi < \phi_2,
	\end{cases}
	\end{equation}
	with $\alpha_1>0$, $\alpha_2<0$, $\alpha_3>0$ and $V_2 = V_1 [ 1 + \alpha_2(\phi_2-\phi_1) ]$ to ensure continuity at $\phi_2$.  A trajectory close to the attractor and starting from large $\phi$ arrives at $\phi_1$ with the velocity $-\alpha_1$, which is chosen to be insufficient to climb the barrier: $\alpha_1^2 \ll (V_2-V_1)/V_1$.  We choose $\bar\phi$ a bit larger than $\phi_1$, and choose $\bar\pi$ in such a way that the field (following a non-attractor trajectory) can pass the barrier.  The resulting PDFs (which do exhibit truncation) for two values of $\sigma_{\delta\phi}$ are presented in Figures~\ref{fig:shutoff-PDF-4} and \ref{fig:shutoff-PDF-5}.  Alongside, the PDFs corresponding to a few smoothed potentials (for which the truncation does not occur) are also shown for comparison.  This smoothing is achieved 
	as follows.  The potential~\eqref{3-segment-V} appears in the equation of motion~\eqref{KG} via $V'/V$, which is itself comprised of three different pieces.  We smooth $V'/V$ by replacing the step functions implicit there by 
	\begin{equation}\label{sharpness}
	\frac{1 \pm \tanh s(\phi-\phi_{1,2})}{2}.
	\end{equation}
	As the sharpness parameter $s$ approaches infinity, the original sharp step functions are recovered.
	
	A few remarks are in order:  In both Figures~\ref{fig:shutoff-PDF-4} and \ref{fig:shutoff-PDF-5} the PDF suddenly truncates at some finite $\delta \N_{\rm max}$ (namely, at $\delta \N_{\rm max}\simeq 0.15$).  So we clearly have truncation.  It is also evident that none of the smoothed barriers lead to complete truncation, although as they become sharper they fall off faster and approach a vertical slope as in the truncated PDF.  The red dashed curve is a PDF obtained from the linear approximation $\delta \N = \N'(\bar\phi) \delta\phi$.  (We have only shown the Gaussian PDF for the sharp barrier $s=\infty$; it is indistinguishable from the Gaussian PDF for $s=10^6$ on all of our plots.)  For $\sigma_{\delta\phi}=10^{-4}$ (Figure~\ref{fig:shutoff-PDF-4}) the right side of the exact PDF for sufficiently sharp peak is skewed to the left compared to the Gaussian PDF.  As a result, even before the truncation $\delta \N_{\rm max}$, we have suppression of $\rho_{_{\delta \N}}$ compared to the linear approximation.  On the contrary, we have skewness to the right for $\sigma_{\delta\phi}=10^{-5}$ (Figure~\ref{fig:shutoff-PDF-5}), which amounts to enhancement of $\rho_{_{\delta \N}}$ before it is completely truncated.  In this case,  a slight smoothing that disallows truncation makes the abundance of PBHs larger than the linear calculation.
	
	It is important to contrast the PDF truncation of this section with that of Sec.~\ref{sec:linear}.  In both cases $\rho_{_{\delta \N}}$ vanishes at some $\delta \N_{\rm max}$.  But while $\rho_{_{\delta \N}}$ of Sec.~\ref{sec:linear} is continuous at $\delta \N_{\rm max}$, here $\rho_{_{\delta \N}}$ is discontinuous at $\delta \N_{\rm max}$ and the truncation occurs suddenly.  We thus name this novel feature a first-order truncation, in analogy with first-order phase transitions. The physical reason behind this difference is the following. Here the height of the step is the same for all trajectories, as long as they start below the step; and it is the existence of the threshold velocity that leads to the truncation. On the contrary, for the linear potential of Sec.~\ref{sec:linear}, each different trajectory needs to climb a different distance to reach the summit, all with the same initial velocity. So, in this case, we have a threshold distance rather than a threshold velocity. To also contrast these two situations with the model studied in Sec.~\ref{sec:quadratic} (i.e., the truncation from the quadratic potential), we note that in the latter situation we encounter a threshold time (or $e$-folds), i.e., the time to reach the end point $\phi_c$ cannot be arbitrarily large. See the discussion below Eq.~\ref{delta-phi-zeta-quadratic}.
	
	As a final remark, note that the stochastic effects can render the PDF nonzero beyond $\delta \N_{\rm max}$ and prevent truncation.  This is because quantum diffusion can introduce stochastic jumps by which the inflaton can pass the barrier and arrive at the classically unreachable point $\phi_c$.  However, we should bear in mind that stochastic effects are known to produce exponential tails~\cite{Ezquiaga:2019ftu}, so although nonzero, they are often (and, in particular, for our choice of parameters) too small to make a significant change. This statement also holds true for the model studied in Sec.~\ref{sec:linear} where we encountered second-order truncation.
	
	\section{Conclusions and outlook}
	\label{sec:conclusions}
	
	In this paper, we carefully analyzed the $\delta \N$ formalism as a non-perturbative method for computing the  PDF of primordial fluctuations. We discussed some subtleties that may arise concerning large fluctuations (which are not expected to appear in the perturbative applications) and argued for a prescription to overcome them. This led to a PDF described in Eq.~\eqref{full_pdf} valid for a broad range of fluctuations and for not necessarily slow-roll inflationary models. Furthermore, we studied several non-attractor single-field inflationary models, computed the full PDF, and showed diversity in the behavior of the PDF for large values of fluctuations. These non-trivial tail behaviors differ significantly from the predictions of the linear perturbation theory. Table \ref{tab:summary} summarizes the tail behavior in different examples we have studied. 
	
	Recently, it has been realized that in models predicting a large abundance of PBHs the loop corrections may dominate over the tree-level contributions spoiling the perturbation theory.  In fact, in many examples that we have studied, ${\cal L}_3[\zeta]/{{\cal L}_2[\zeta]}$ is pretty large for $\zeta\simeq 1$ (as opposed to ${\cal L}_3[\delta \phi]/{{\cal L}_2[\delta \phi]}$ which remains small) so that the non-perturbative treatment is indeed necessary. However, note that the loop-level contributions to the statistics of $\zeta$ are already taken care of in our formalism since we treated the relation between $\delta \phi$ and $\zeta$ non-perturbatively. Therefore, these non-perturbative effects (discussed e.g., in \cite{Inomata:2022yte}) are not concerning. On the other hand, since $\zeta$ must grow significantly in the non-attractor phase to account for the PBH abundance, loop corrections on the CMB scale modes due to these enhanced short-scale fluctuations may dominate over the tree-level amplitude \cite{Kristiano:2022maq, Riotto:2023hoz, Firouzjahi:2023aum}. While this claim remains debatable in the explored setups and depends on the details of the transition to the slow-roll regime, we note that the different potentials studied here may open up new ways to overcome this possible difficulty.  Whether avoiding these effects can be easily achieved or it requires fine-tuning is an interesting open question that deserves careful analysis, beyond the scope of this paper.  At the very least, since the tail of the distribution is so non-trivial for our examples, one might be able to produce a large abundance of PBHs with smaller values of power spectrum (i.e., a smaller power spectrum may be compensated by a heavier-than-Gaussian tail). A reduced short scale power spectrum then downgrades the effect  of the loop corrections on the CMB scale power spectrum. As a related remark, notice that taking into account the effect of the tail on the PBH abundance may also alter the amount of fine-tuning required in the parameters of the model to produce a fair amount of PBHs which is another, worth exploring, open question \cite{Cole:2023wyx}. 
	
	Finally, we stress that extending our formalism to the case of multiple-field models of inflation is straightforward. We expect the multiple-field setup to show even more diverse possibilities~\cite{Panagopoulos:2019ail,Hooshangi:2022lao,Achucarro:2021pdh}. Furthermore, the multiple-field setup may offer easier ways to relax the fine-tunings and to avoid the large loop effects mentioned above. Therefore, exploring multiple-field models of inflation are, in various ways, well-motivated and can be considered as the next move.

	\begin{table}[]
		\center 
		\label{tab:summary}
		\caption{Summary of the tail behaviors for the examples studied in this paper.  The various constants appearing in some equations are collectively named $c$ or $\tilde c$; refer to the cited section to find the details. We always assume that $\phi_c>\bar \phi$, i.e., the net motion of the field is from large values to small values.}
		\begin{tabular}{|l|l|l|}
			\hline
			Section & $V(\phi)$                                                & Feature of the PDF at large $\zeta$                                    \\ \hline\hline
			\ref{sec:linear}  & $V_0 \left( 1+ \alpha \phi \right)$; $\alpha>0$ & Gaussian tail but not identical to the linear theory   \\ \hline
			\ref{sec:linear}  & $V_0 \left( 1+ \alpha \phi \right)$; $\alpha<0$ & 2nd-order truncation   \\ \hline
			\ref{sec:quadratic}  & $V_0 \left( 1+ \frac{1}{2}\beta \phi^2\right)$; $\beta<\frac34$ & Double exponential tail $(\exp(- c\, e^{c_2 \zeta}))$ \\ \hline
			\ref{sec:quadratic}  & $V_0 \left( 1+ \frac{1}{2}\beta \phi^2\right)$ ; $\beta>\frac34$ &  Two peaks, 2nd-order truncation  \\ \hline
			\ref{sec:power-law}   & $V_0(1+c \, \phi^{2+1/p}) $                                         & Power law tail   $(\tilde c/\zeta^{p+1})$                                  \\ \hline
			\ref{sec:barrier} & Sharp step in $V(\phi)$ 				      & 1st-order truncation                                    \\ \hline
		\end{tabular}
	\end{table}

	\begin{acknowledgments}
		M.N.\ acknowledges financial support from the research council of University of Tehran.
	\end{acknowledgments}
	
	\appendix
	
	\section{More on the power-law tail}
	\label{app:V-p}
	
	As we mentioned before, the analysis in Sec.~\ref{sec:power-law} was only suggestive and not a rigorous derivation of the potential~\eqref{V-p}.  We now take a different look at this potential and provide more convincing evidence that $\delta\N$ indeed satisfies Eq.~\eqref{deltaN-deltaphi}, thereby justifying the power-law tail.  
	
	\begin{figure}
		\centering
		\includegraphics[scale=.5]{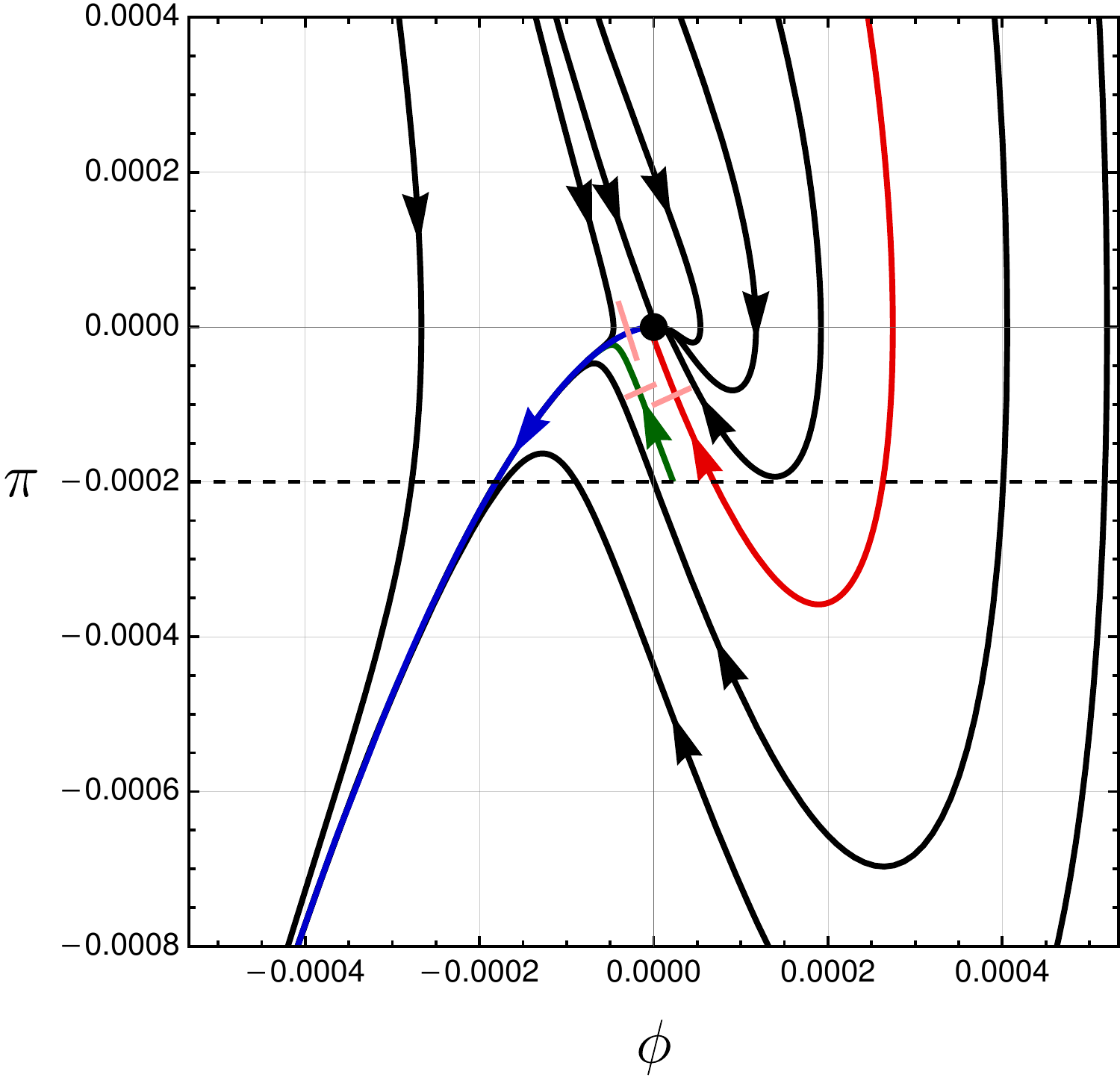}
		\caption{The phase portrait of the dynamical system~\eqref{dyn-sys} corresponding to motion under the potential~\eqref{V-p}.  The red and blue curves are, respectively, incoming to and outgoing from the fixed point, while the tiny segments on them indicate distances $\varepsilon_1$ and $\varepsilon_2$ from the origin.  The trajectory of interest (green) starts with initial velocity $\bar\pi$ from $\bar\phi+\delta\phi$, goes up close to $\bar\phi=0$ spending a long time in the nearby bottleneck area, and then turns down toward $\phi_c$ (not shown in the figure).  The tiny segment splits the green curve into a right piece (close to the red curve) and a left piece (close to the blue curve).  }
		\label{fig:phase-V-p}
	\end{figure}
	
	Let us study the equation of motion~\eqref{KG} with the potential~\eqref{V-p} in the phase space.  It amounts to
	\begin{equation}\label{dyn-sys}
	\begin{aligned}
	\phi' &= \pi, \\
	\pi' &= - \left( 3 - \frac12 \pi^2 \right) \left( \pi + p\gamma^{1/p} \phi^{1+1/p} \right),
	\end{aligned}
	\end{equation}
	where we have set $\bar\phi=0$ (without loss of generality) and used the approximation $\gamma|\phi-\bar\phi|^{p+1}\ll1$ (de Sitter limit).  The corresponding phase portrait is depicted in Figure~\ref{fig:phase-V-p} where a few trajectories are shown.  {Eventually our focus will be the green trajectory, but let us first explain a few key features of the phase portrait.}  The point $(\phi=0, \pi=0)$ at the origin is a fixed point of this system: if you start from $\phi=0$, with zero initial velocity, you will stay there forever.  Two notable trajectories are shown too: The red curve that ends up at the origin, and the blue curve that starts off from the origin.  The red curve is the trajectory of an inflaton that rolls on the flat plateau on $\phi>0$ with an initial velocity fine tuned such that at time $N=\infty$ it reaches and stops at $\phi=0$;  any trajectory with a faster velocity will overshoot $\phi=0$.  It can be shown that in the vicinity of the origin, where the dynamics is like USR, the red curve asymptotes to $3\phi+\pi=0$.  Furthermore, the time along the red curve to reach a distance $\varepsilon_1$ from the origin, grows like
	\begin{equation}\label{divN-red}
	\int^{\varepsilon_1} \frac{-d\phi}{\phi'} \approx \int^\delta \frac{d\phi}{3\phi} \sim \log \varepsilon_1.
	\end{equation}
	On the other hand, the blue curve is the trajectory of an inflaton that starts at rest from the origin at time $N=-\infty$.  In the neighborhood of the origin, the blue curve can be approximated by $\pi + p\gamma^{1/p} \phi^{1+1/p} = 0$,\footnote{To derive the equation of the blue attractor curve, assume that in the neighborhood of the origin it is given by $\phi(n) = cn^{-k}$, where $k$ is a positive constant so that it reaches $\phi=0$ at $n=-\infty$.  Notice that this is part of the definition of the attractor, since all the other neighboring curves diverge away from $\phi=0$ at $n=-\infty$.  Then plug this in the first line of Eq.~\eqref{dyn-sys} to obtain $\pi(n) = -ckn^{-k-1}$; and then in its second line to obtain
		\begin{equation}
		ck(k+1)n^{-k-2} = 3ckn^{-k-1} - 3p\gamma^{1/p} \left( cn^{-k} \right)^{1+1/p},
		\end{equation}
		where we have ignored $\frac12\pi^2$ next to $3$.  Now owing to the large $n$ limit, we must neglect the $n^{-k-2}$ term compared with the $n^{-k-1}$ term.  Therefore, at this order of the approximation, both sides of the preceding equation must equal zero.  The vanishing of the right hand side gives the claimed equation of the blue curve.  In fact, it also determines the constants  to be $c=1/\gamma$ and $k=p$, but we don't need them here.  It is to be emphasized that this equation governs the blue curve only in a small neighborhood of the origin and the true equation has higher terms that are suppressed by negative powers of $n$.  Finally, note that the same approach could be applied to obtain the equation of the red curve in the USR stage, albeit using $\phi(n) = e^{-an}$ instead of the power-law ansatz, which yields $a=3$. }
	and the time along it starting from a distance $\varepsilon_2$ from the origin, behaves like
	\begin{equation}\label{divN-blue}
	\int_{-\varepsilon_2} \frac{-d\phi}{\phi'} \approx \int_{-\varepsilon_2} \frac{d\phi}{p\gamma^{1/p} \phi^{1+1/p}} \sim \left( \gamma \varepsilon_2 \right)^{-1/p}.
	\end{equation}
	It is clear from Eqs.\eqref{divN-red} and ~\eqref{divN-blue}   that the time $N$ diverges along both the red and blue curves as they approach the fixed point, i.e., as $\varepsilon_1, \varepsilon_2 \to0$.
	
	Let us now focus on a trajectory in the phase space (like the green curve in Figure~\ref{fig:phase-V-p}) that passes close by the origin (corresponding to the large values of $\delta \N$, which are of interest for exploring the tail of PDF).   This trajectory starts from $(\bar\phi+\delta\phi,\bar\pi)$ where it is close to the red curve (since the onset of the motion is in the USR stage), goes up on the phase plane close to the origin, and then turns around and approaches the blue curve (which is our final slow-roll attractor stage).  Thus it should be clear that the trajectories of interest have two asymptotic behaviors: approaching the red curve in the past, and approaching the blue curve in the future.  Since we work with fixed initial velocity $\bar\pi$, our control parameter is $\varepsilon = \delta\phi_{\rm max} - \delta\phi$ which measures how close to the red curve we start: the smaller $\varepsilon$, the closer we get to the origin.  As the trajectory is moved closer to the origin ($\varepsilon\to0$), the time $\N$ it takes to reach $\phi_c$ becomes longer.  In other words, the region around the origin acts like a bottleneck that requires a very long time to traverse.  This is because of the infinite time it takes to reach/depart from the origin on the red/blue curves.  We can evaluate $\N$ by breaking the green curve up into two pieces, and estimate each piece by a segment of the red/blue curve that goes as close to the origin as $\varepsilon_1$/$\varepsilon_2$, with both $\varepsilon_1$ and $\varepsilon_2$ proportional to $\varepsilon$.  Therefore, according to Eqs.~\eqref{divN-blue} and \eqref{divN-red}, the major contributions to $\N$ along the green trajectory are given by the two terms $\sim \log\varepsilon$ and $\sim (\gamma\varepsilon)^{-1/p}$, of which the latter is the dominant; thus
	\begin{equation}
	\N \sim \left\{ \gamma \left( \delta\phi_{\rm max} - \delta\phi \right) \right\}^{1/p}.
	\end{equation}
	Since the difference between $\N$ and $\delta\N$ is finite, this is precisely the behavior we had claimed in Eq.~\eqref{deltaN-deltaphi}.
	
	\bibliography{examples}
	\bibliographystyle{JHEP}
	
\end{document}